\documentclass[twocolumn]{aastex6}

\usepackage{mathptmx}
\usepackage[T1]{fontenc}
\usepackage{ae,aecompl}
\usepackage{graphicx}	
\usepackage{amsmath}	
\usepackage{amssymb}
\usepackage{tabularx}
\usepackage{bm}
\usepackage{morefloats}

\def\feh{\mathrm{[Fe/H]}}
\def\afe{[\alpha/\mathrm{Fe}]}
\def\dex{\mathrm{dex}}
\def\age{\mathrm{Age}}
\def\vlos{v_\mathrm{los}}
\def\disp{\sigma_\mathrm{int}}

\def\kms{\mathrm{km\ s^{-1}}}

\shorttitle{Magellan/M2FS A267}
\shortauthors{Tucker et. al.}

\begin{document}

\title{Magellan/M2FS Spectroscopy of Galaxy Clusters: Stellar Population Model and Application to Abell 267}

\author{Evan Tucker}
\author{Matthew G. Walker}
\affil{McWilliams Center for Cosmology, Carnegie Mellon University, Pittsburgh, PA, USA}
\author{Mario Mateo}
\affil{Department of Astronomy, University of Michigan, Ann Arbor, MI 48109-1042}
\author{Edward W. Olszewski}
\affil{Steward Observatory, University of Arizona, Tucson, AZ 85721}
\author{John I. Bailey III}
\affil{Leiden Observatory, Leiden University, P.O. Box 9513, 2300RA Leiden, The Netherlands}
\author{Jeffrey D. Crane}
\and
\author{Stephen A. Shectman}
\affil{Carnegie Observatories, 813 Santa Barbara Street, Pasadena, CA 91101, USA}

\begin{abstract}

We report the results of a pilot program to use the Magellan/M2FS spectrograph to survey the galactic populations and internal kinematics of galaxy clusters.
For this initial study, we present spectroscopic measurements for $223$ quiescent galaxies observed along the line of sight to the galaxy cluster Abell 267 ($z\sim0.23$).
We develop a Bayesian method for modeling the integrated light from each galaxy as a simple stellar population, with free parameters that specify redshift ($\vlos/c$) and characteristic age, metallicity ($\feh$), alpha-abundance ($\afe$), and internal velocity dispersion ($\disp$) for individual galaxies.
Parameter estimates derived from our 1.5-hour observation of A267 have median random errors of $\sigma_{v_\mathrm{los}}=20\ \mathrm{km\ s^{-1}}$, $\sigma_\age=1.2\ \mathrm{Gyr}$, $\sigma_\feh=0.11\ \mathrm{dex}$, $\sigma_{[\alpha/\mathrm{Fe}]}=0.07\ \mathrm{dex}$, and $\sigma_{\sigma_\mathrm{int}}=20\ \mathrm{km\ s^{-1}}$.
In a companion paper, we use these results to model the structure and internal kinematics of A267.
\end{abstract}

\keywords{galaxies: clusters: individual(Abell 267) -- techniques: imaging spectroscopy -- galaxies: general}

\section{Introduction}
\label{intro}

Galaxy clusters are the most massive gravitationally bound and collapsed structures in the Universe, and therefore they are important laboratories for observational cosmology \citep[][]{Diaferio1997, Dressler2004, Voit2005, Jones2009, Vikhlinin2009, Rines2013, Geller2013}.
Due to their high density of galaxies they are also ideal for studying galaxy interactions and the effect these interactions have on the galaxy population.
Galaxy clusters are studied in a multitude of ways, from gravitational lensing, both weak and strong \citep[for example][and references therein]{Kneib2008, Applegate2014, Barreira2015, Gonzalez2015}  to X-ray temperature measurements of hot intracluster gas \citep[][]{Guennou2014, Moffat2014, Girardi2016, Rabitz2017} to Sunyaev-Zeldovich effects \citep[][]{Sunyaev1970, Churazov2015} to optical spectroscopy \citep[e.g.][and references therein]{Rines2003, Rines2013, Hwang2014, Stock2015, Tasca2016, Biviano2016, Dressler2016, Sohn2017}.
Many of these methods seek to measure the mass and/or the mass function of the cluster, thus constraining cosmological parameters such as the amplitude of the power spectrum or the evolution of matter and dark energy densities over cosmological time.

With the advancement of multi-object spectrographs, astronomers have the ability to conduct large spectroscopic surveys of galaxies in cluster environments.
Multiple-object spectroscopic systems have allowed for observations of hundreds of objects simultaneously.
These spectrographs provide the necessary tools to perform efficient follow up of photometrically identified galaxies over a range of redshifts.
For example, the Sloan Digital Sky Survey (SDSS) produced a spectroscopic catalog of millions of galaxy spectra with up to a thousand cluster member galaxies at low redshift and less than ten member galaxies at their highest redshift $z\sim0.8$ \citep[][]{BOSS2012, SDSS2016}.
Additionally, the new age of spectroscopic data from SDSS includes integral field unit (IFU) observations with MApping Nearby Galaxies at Apache Point Observatory (MANGA) which resolves galaxy spectra in two-dimensions on the sky.
Using the 6.5m MMT and Hectospec fiber spectrograph, \citet{Rines2013} have measured redshifts for more than 22,000 individual galaxies in 58 clusters (the HECs survey).
Moreover, astronomers have used MMT/Hectospec and VLT/VIMOS to build large spectroscopic catalogs for cluster galaxies observed with the Cluster Lensing and Supernova Survey with Hubble (CLASH) \citep[e.g.][]{Geller2014, Biviano2013, Rosati2014, Girardi2015}.
Another commonly used spectrograph, The Inamori-Magellan Areal Camera and Spectrograph (IMACS) is a multi-slit, wide-field spectrograph on the Baade-Magellan Telescope in Chile, which has been used in recent years to study galaxy clusters \citep{Dressler2011, Oemler2013}.

The Michigan/Magellan Fiber System (M2FS) is a multi-object fiber spectrograph consisting of 256 fibers and was installed on the 6.5m Clay-Magellan Telescope at the Las Campanas Observatory in Chile in August 2013 \citep[][]{Mateo2012, Bailey2014}.
In its highest resolution setting ($R\sim50000$), M2FS has been used by \citet{Bailey2016} to search for exoplanets in open clusters, \citet{Johnson2015a} to measure chemical abundances in globular clusters \citep[see also][]{Johnson2015b, Roederer2016a, Johnson2017}, and \citet{Roederer2016b} to measure chemical abundances in dwarf spheroidal galaxies.
At more moderate resolutions ($R\sim18000$) \citet{Walker2015b,Walker2016} and \citet{Simon2015} have used M2FS for detailed kinematic analyses of dwarf spheroidal galaxies.

In addition to cosmological constraints from cluster masses, the galaxy spectra themselves convey a multitude of information about their stellar populations.
In recent years, with the development of more robust statistical techniques, there has been great progress in the fitting of galaxy spectra to extract stellar population information.
These efforts have focused on building a more robust statistical framework around the early methods of stellar population synthesis \citep[][]{Tinsley1972, Searle1973, Larson1978} used for modeling the spectral energy distributions (SEDs) of galaxies.
These early stellar population synthesis methods have been improved over the years to incorporate a more complete understanding of galactic processes \citep[see][for a review]{Walcher2011}.
In the past few years, new efforts have been made to apply Bayesian techniques to fit these stellar population models.
BayeSED \citep[][]{Han2014} and \textsc{beagle} \citep[][]{ Chevallard2016} are two recently developed Bayesian models aimed at fitting SEDs of galaxies over a large wavelength coverage.
However, these models are geared towards SEDs, which sample only at few band passes over a large wavelength range (from $\gamma$-rays to IR).
And most recently, \citet{Meneses2015} developed a single stellar population model with Bayesian statistical techniques to fit spectra in the near-infrared.

In this paper we develop an integrated light population synthesis method for fitting galaxy spectra built upon the modeling techniques developed by \citet{Walker2015a}.
We applied this method to spectra obtained in November 2013 of Abell 267 (A267) with the Michigan/Magellan Fiber System (M2FS) on the Clay-Magellan Telescope at Las Campanas Observatory in Chile.
In \S\ref{observations} we describe the observations and subsequent data reduction.
\S\ref{ILSSPM} describes the integrated light spectral model used to fit these spectra.
\S\ref{Analysis of Spectra} we describe how we implemented and fit this model and test it with mock spectra and previously fit galaxy spectra.
And finally in \S\ref{results} we apply the model to fit the new A267 spectra.

\section{Observations and Data Reduction}
\label{observations}
In this section we present a pilot program for cluster spectroscopy at low resolution with M2FS and detail the reduction of these spectra.
\subsection{Target Selection}
\label{TargetSelection}
\begin{figure*}
\centering
\includegraphics[width=\textwidth]{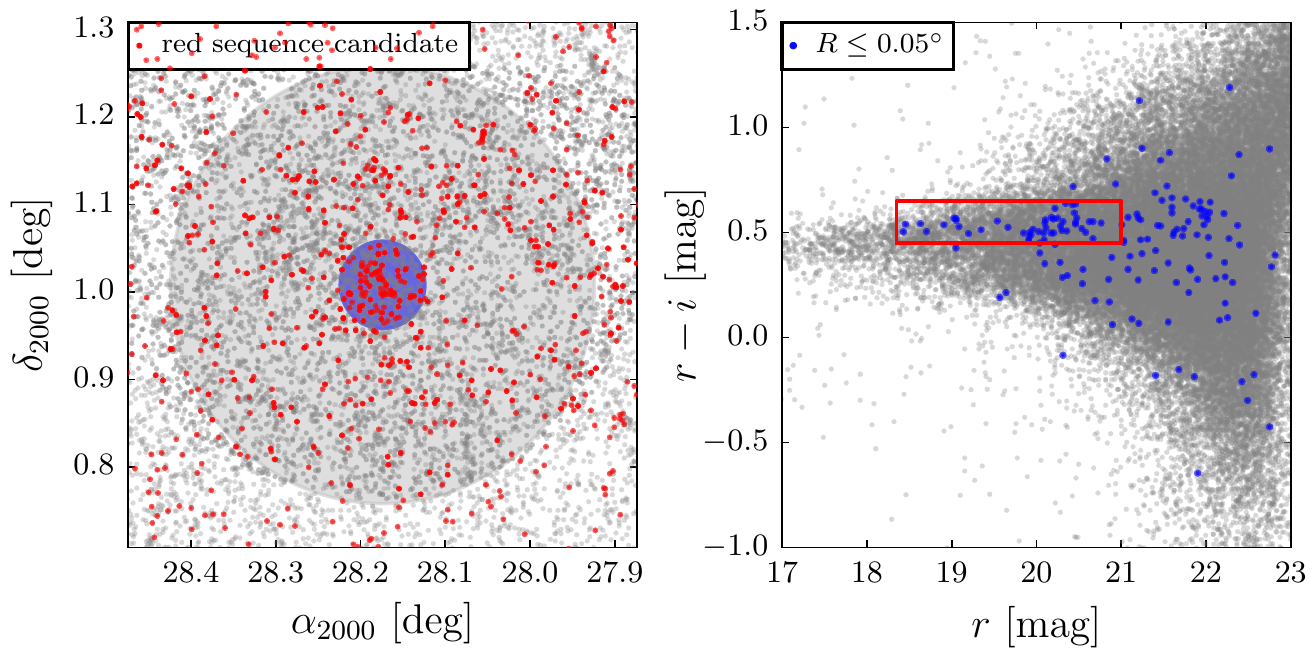}
\caption{Equatorial coordinates (left) and $r$, $r-i$ photometry (right) for galaxies along the line of sight to Abell 267 \citep[SDSS DR12,][]{SDSSDR12}.  In the right-hand panel, blue markers represent galaxies nearest the center of Abell 267 (from within the shaded blue circle in the left-hand panel).  The red rectangle encloses quiescent galaxies on Abell 267's red sequence.  In the left-hand panel, red points show the spatial distribution of these red sequence candidates; it is from this set of objects that we select M2FS targets.  In the left-hand panel, the shaded gray circle represent the M2FS field of view.}
\label{A267_Targets}
\end{figure*}

We select targets for M2FS observations by identifying galaxies detected in Sloan Digital Sky Survey images \citep[Data Release 12,][]{SDSSDR12} that are projected along the line of sight to Abell 267 and are likely to be quiescent cluster members.
First we extract from SDSS all extended sources projected within a circle of diameter $0.5^\circ$ that is centered on Abell 267 ($\alpha_{2000}=28.174^{\circ}$ , $\delta_{2000}=+1.008^{\circ}$); for all such objects brighter than r=23, Fig. \ref{A267_Targets} displays sky positions and r, r-i photometry.
In the right panel of Fig. \ref{A267_Targets}, blue markers indicate colors and magnitudes for galaxies nearest the center of Abell 267---i.e., those lying within the shaded blue circle (radius $0.05^\circ$) in the left panel of Fig. \ref{A267_Targets}.
These objects clearly trace A267's red sequence, which is enclosed by a red rectangle in the right panel of Fig. \ref{A267_Targets}.
Finally, red points in the left panel of Fig. \ref{A267_Targets} indicate sky positions for all galaxies lying in the red sequence selection box.
We consider all objects within this selection box to be candidate members of Abell 267's red sequence.
It is from this set of objects that we select M2FS targets, giving greater weight to brighter objects.

\subsection{Observations}
We observed 223 individual galaxy spectra on 30 November 2013 on the Clay Magellan Telescope using M2FS.
We used the low resolution grating on M2FS and chose a coverage range of 4600-6400 \AA\ with a resolution of $R\sim2000$.
Of the 256 fibers available on M2FS, we allocated 223 for science targets, leaving 33 fibers directed at relatively blank regions of sky.
We observed the field over 6 sub-exposures of 15 minutes each, which we then stacked to improve signal-to-noise ratio and remove cosmic rays (see \S\ref{data reduction} below).
For wavelength calibration we took Thorium-Argon-Neon lamp exposures both before and after a set of science exposures, and we took a quartz-lamp exposure immediately after the sequence of science exposures.
To identify the apertures on the CCD for each fiber we took ``fibermap" exposures which are high signal-to-noise (S/N) exposures of the ambient light in the dome during the daytime.
For the purpose of calibration and correction for variations in fibre throughput, we also took a series of high S/N exposures (including Th-Ar-Ne and quartz calibrations) during evening twilight sky.

\subsection{Data Reduction}
\label{data reduction}
\begin{figure}
\centering
\includegraphics[width=\columnwidth]{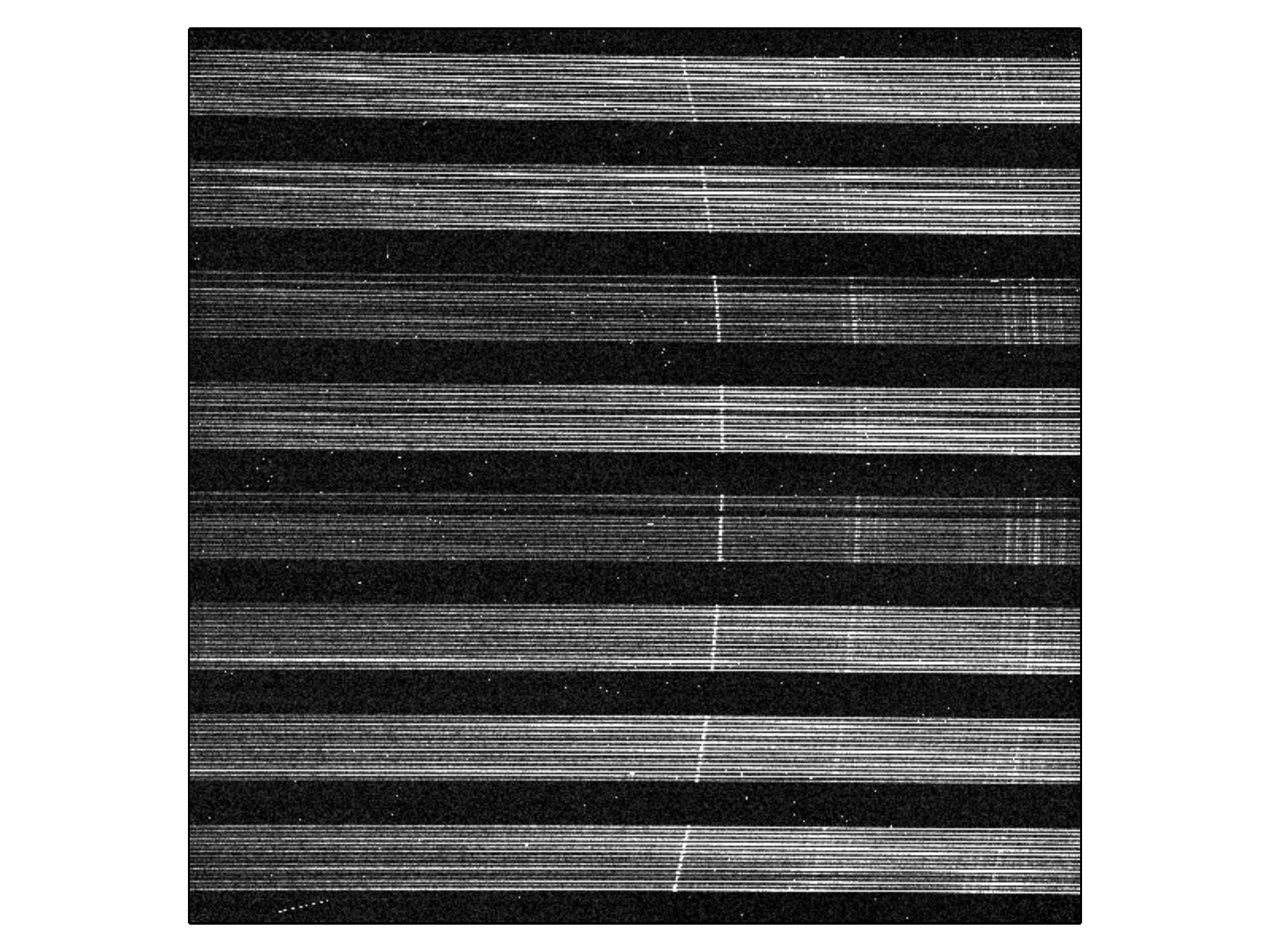}
\caption{One of the two CCDs from M2FS.  Each horizontal line corresponds to one of the fiber spectra.  The fibers are organized into 8 cassettes (fiber bundles) of 16 fibers.  The bright curved vertical feature is a bright atmospheric emission line that is observed in nearly all fibers.}
\label{ccd chip}
\end{figure}

The detector used with M2FS consists of two 4096 x 4112-pixel CCDs, each of which is read out through four amplifiers.
We used the 2 x 2 binning setup for readout, so the output images are 2048 x 2056-pixels.
The 256 fibers are organized into 16 cassettes of 16 fibers each.
The cassettes are spatially separated on the CCD and within each cassette each individual fiber is spatially separated.
Fig. \ref{ccd chip} shows an example of one of the CCDs with twilight spectra obtained during the A267 observations.

We use standard \textsc{iraf} routines to process the raw images, to extract the 1D spectra and to estimate the wavelength solution for each spectrum obtained in each science exposure.
We also propagate the variance associated with the count level in each pixel of each image.
At the outset, for every science frame (i.e. the images obtained in an individual science exposure) we generate a corresponding variance frame in which the value assigned to a given pixel is 
\begin{equation}
\label{pixel variance}
\mathrm{Var(pix)}=C(\mathrm{pix})G+R^2
\end{equation}
where $C(\mathrm{pix})$ is the count in analog-to-digital units (ADU), $G\approx0.68\mathrm{e}^-/\mathrm{ADU}$ is the gain of the M2FS detector and $R=2.7\mathrm{e}^-$ is the read noise.
In order to propagate variances, we process variance frames accordingly to the way that we process their corresponding science frames (see below).
For example, where we combine spectra via addition or subtraction (e.g. to combine subexposures or to subtract sky background) we compute the combined variances as the sum of the variances associated with the pixels contributing to the sum or difference.
Or, where we rescale count levels in a given science exposure (e.g. to correct for the variability in the fiber throughput) we rescale the variances by the square of the same factor.

For a given frame we begin the data reduction pipeline using the \textsc{iraf} package CCDPROC to perform overscan corrections independently for each amplifier.
We then rescale the counts in each frame by the gain associated with each amplifier independently in order to convert ADUs to electrons.
For each of the two CCDs, we combine the four amplifier images to form a continuous gain-corrected image.
We then bias subtract and remove the dark current.
For the dark current correction, we rescale the measured dark current by the exposure time of each individual subexposure, then subtract this rescaled dark current.
During our observations of A267, there was a non-negligible dark current that builds up in the corners of the CCD and contributed $\sim50-200$ counts per 15 minute exposure.

Next, we use the \textsc{iraf} package APALL to identify the locations and shapes of the spectral apertures, and to extract 1D spectra for science, quartz, Th-Ar-Ne arcs, and twilight exposures and associated variance frames.
We initially identify aperture locations and trace patterns in the relatively bright fibermap exposures.
Fibermap exposures are obtained by taking short exposures, with all fibers plugged, of ambient sunlight in the dome during the daytime.
We use fibermaps instead of quartz calibration frames to identify aperture locations because the ambient sunlight more uniformly illuminates all fibers compared to quartz exposures.
After identifying the aperture locations with the fibermaps, we use the \textsc{iraf} package APSCATTER to fit the scattered light in the regions of the CCD outside the apertures and subtract this fit from the regions of the CCD inside the apertures.
Fixing the relative locations and shapes of the apertures according to the fibermaps, we use APALL and allow the entire aperture pattern to shift globally in order to provide the best match to the corresponding science frames.
We apply exactly the same shift to define apertures and traces for the Th-Ar-Ne frames.
We then use APALL to extract the spectra from each aperture by combining (adding) counts from pixels along the axis perpendicular to the dispersion direction for each science, twilight, and Th-Ar-Ne and associated variance frames.

Next we use the extracted twilight spectra to adjust for differences in fiber throughput and pixel sensitivity.
We first fit a (6th-order) Legendre function to the extracted twilight spectra, which iteratively rejects counts that either exceed the fit by more than 3-times the rms of the residuals or are smaller than the fit by more that 1.75-times the rms of residuals.
The lower tolerance is smaller than the upper tolerance to effectively exclude the absorption features from the fit.
We then determine the median count level of the fit for each fiber and normalize each fit by the mean of these median count levels.
Finally we divide the science and twilight spectra by this normalized fit per spectrum, thereby correcting for differences in throughput and pixel sensitivity simultaneously.

Next we estimate wavelength solutions, $\lambda$(pix).
For each extracted Th-Ar-Ne spectrum, we use the \textsc{iraf} package IDENTIFY to fit a (5th-order) Legendre polynomial to the centroids of between 30 and 40 identified emission lines of known wavelength.
Residuals of these fits typically have a root mean square (rms) scatter $\sim0.150$ \AA\ or $\sim10\ \mathrm{km/s}$.
We assign the same aperture-dependent wavelength solutions to the corresponding science frames.
Except for extraction from 2D to 1D, each individual spectrum retains the same sampling native to the detector; therefore, the wavelength solutions generally differ from one spectrum to another and have a non-uniform $\Delta\lambda/\Delta\mathrm{pix}$ even within the same spectrum.

\subsection{Sky Subtraction}
\label{sky subtraction}

\begin{figure*}
\centering
\includegraphics[width=.49\textwidth]{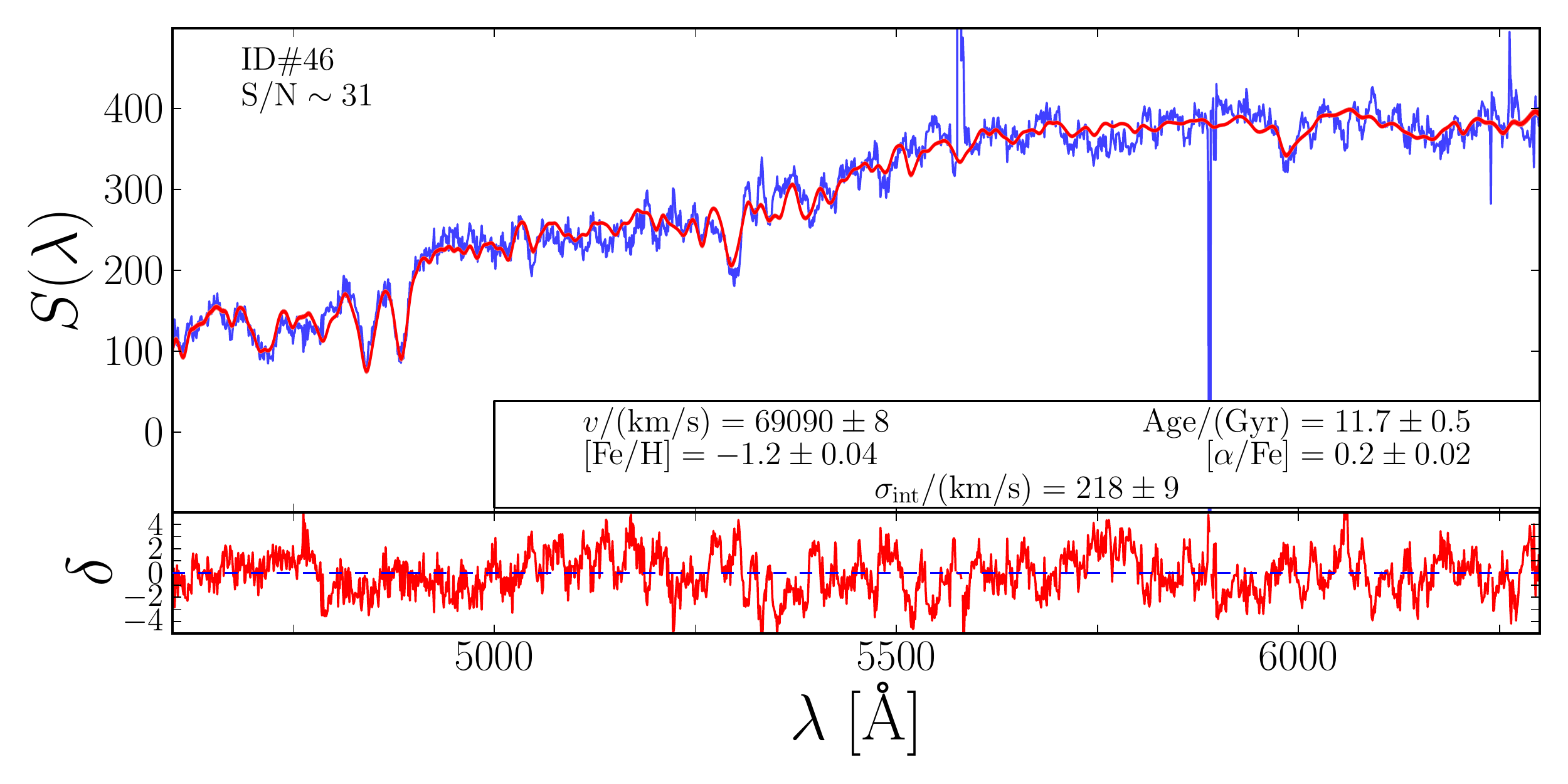}
\includegraphics[width=.49\textwidth]{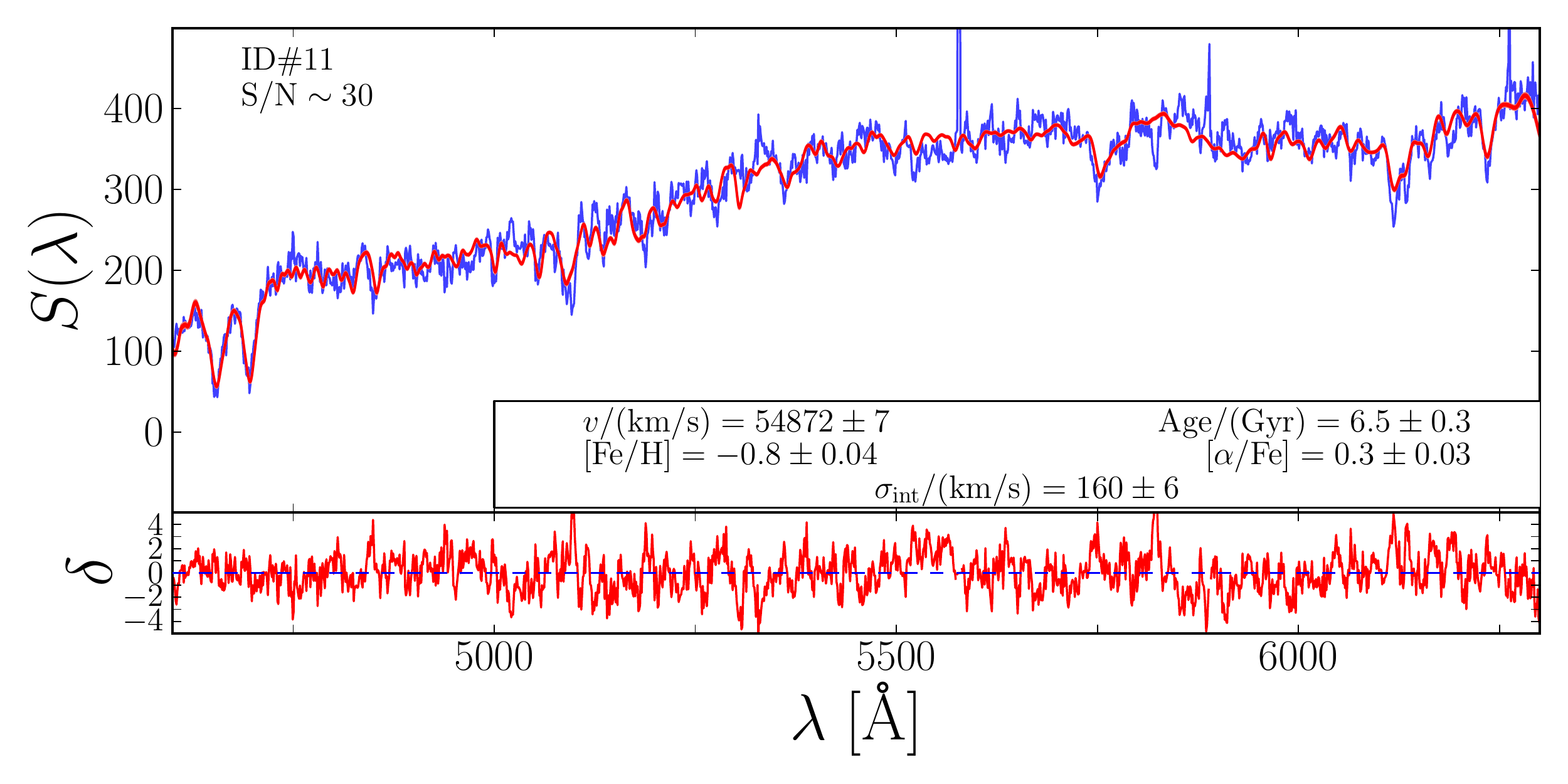}\\
\includegraphics[width=.49\textwidth]{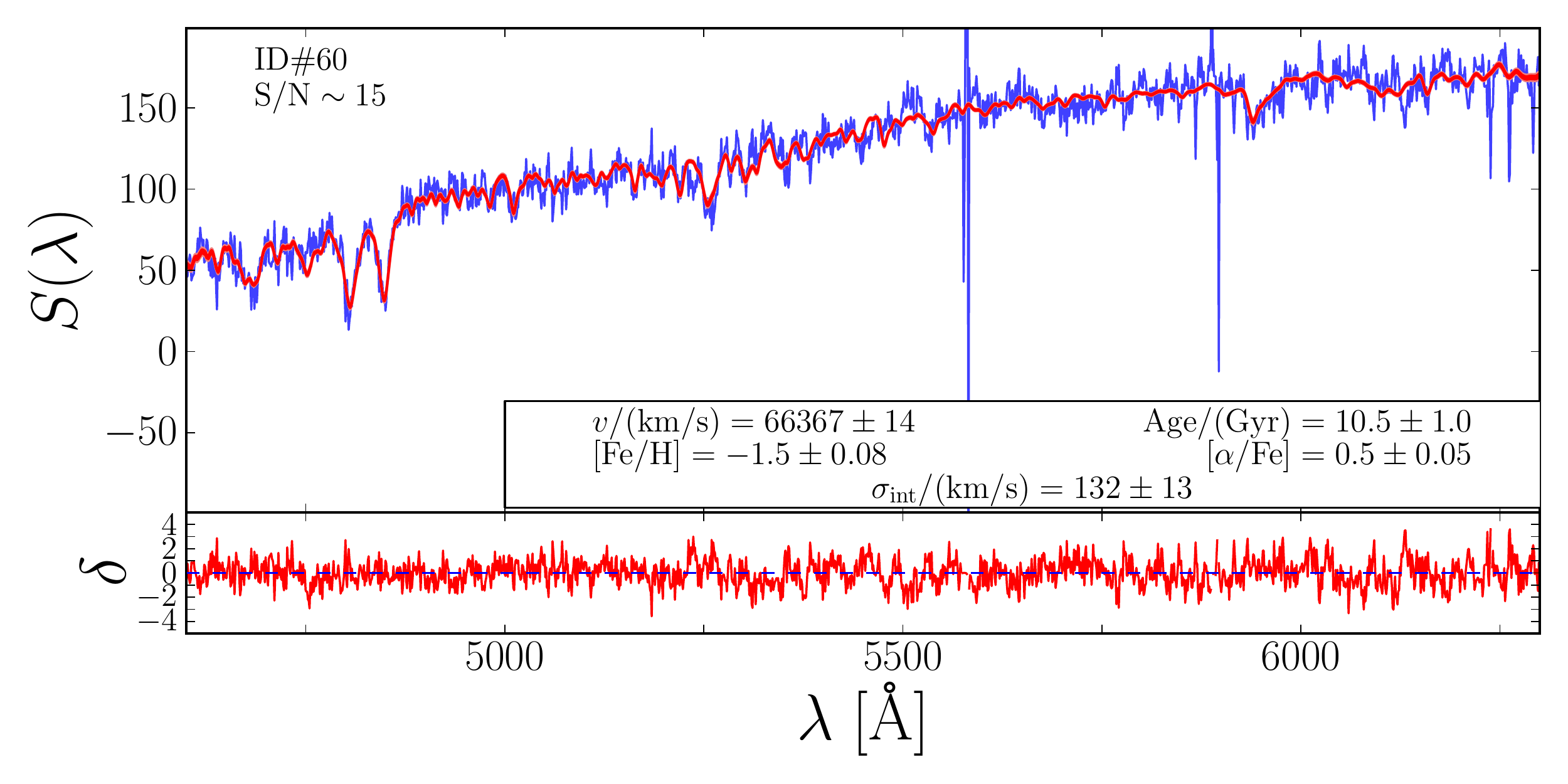}
\includegraphics[width=.49\textwidth]{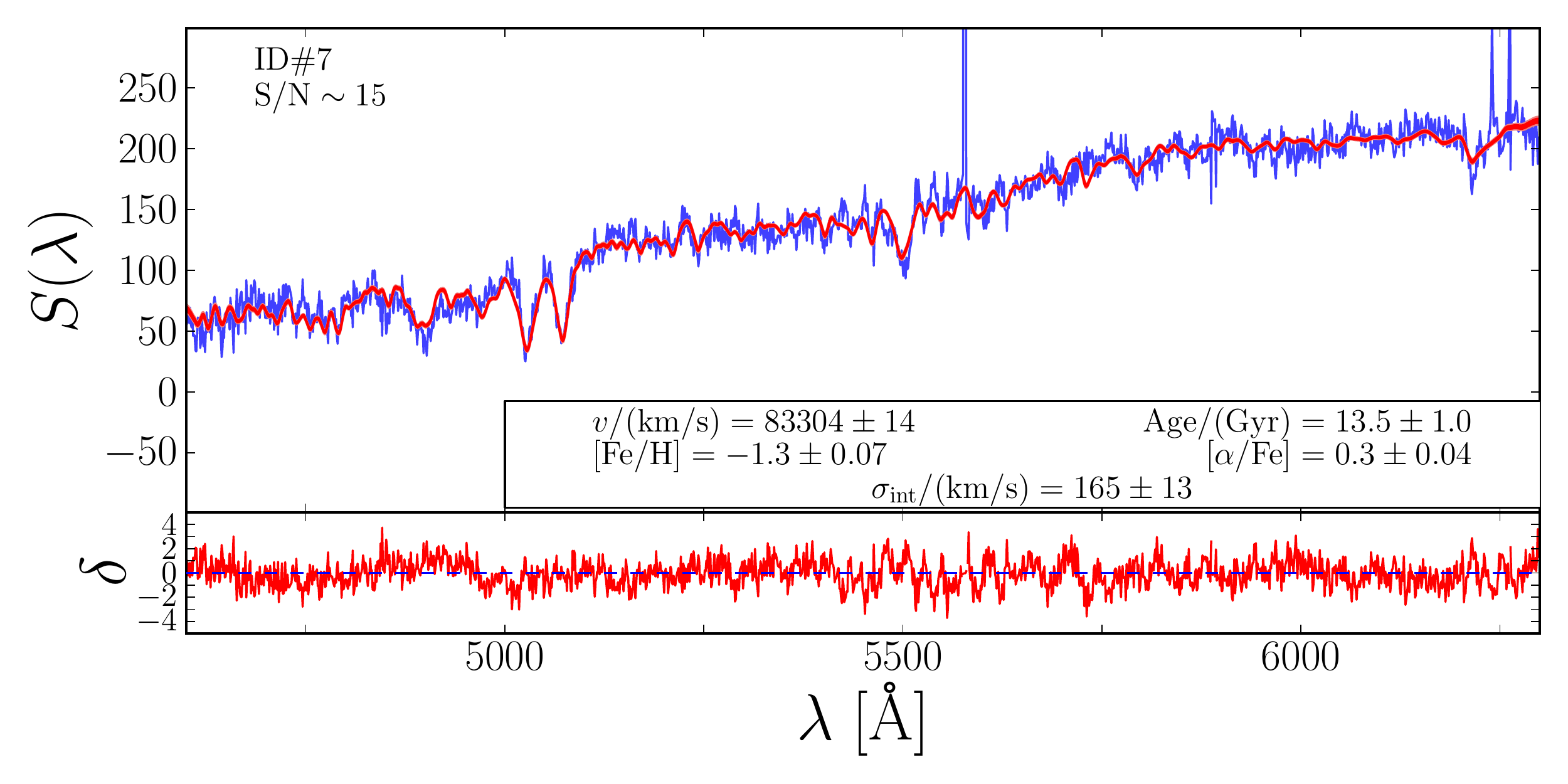}\\
\includegraphics[width=.49\textwidth]{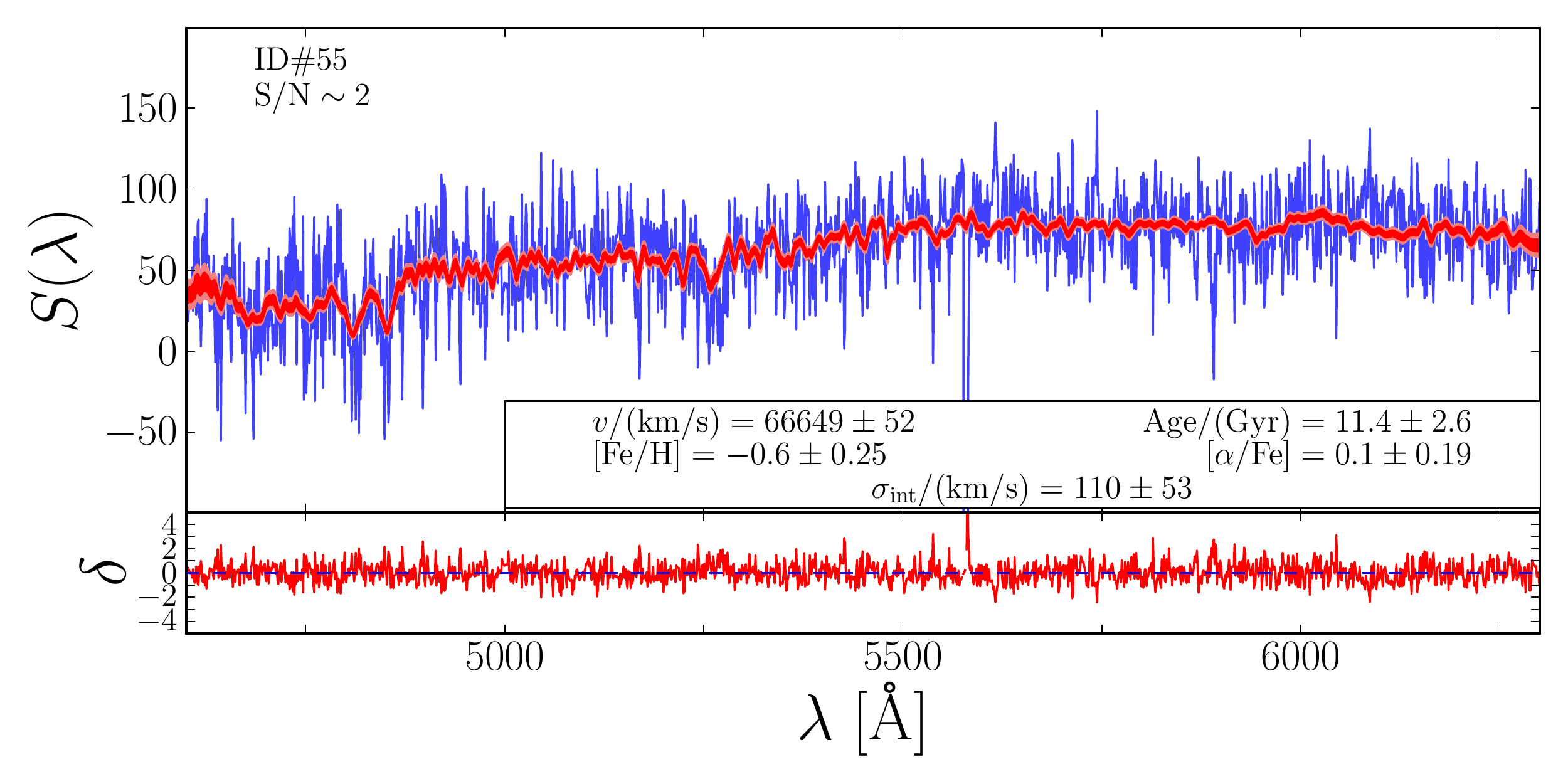}
\includegraphics[width=.49\textwidth]{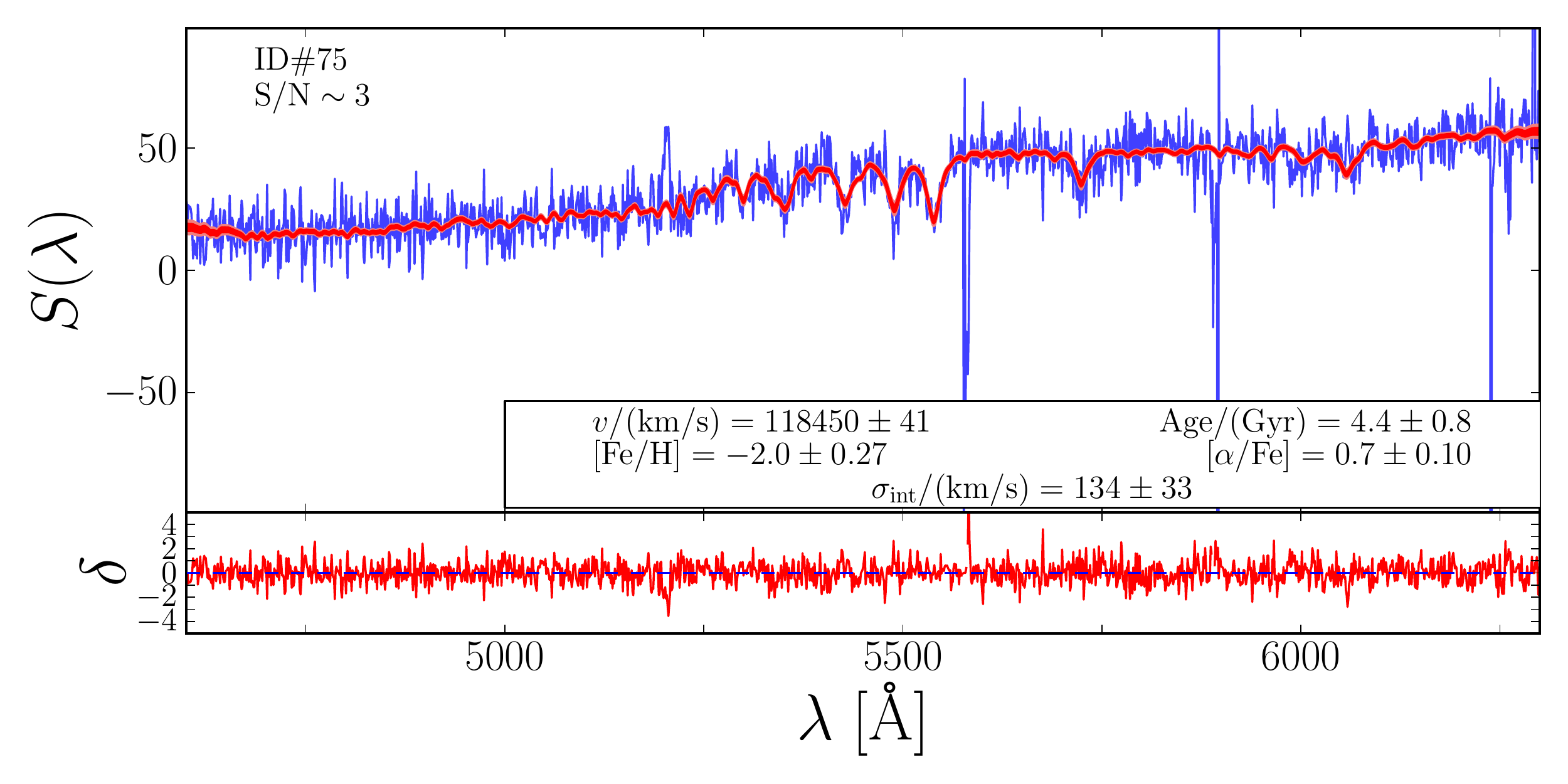}
\caption{Sky-subtracted M2FS spectra (blue) for probable Abell 267 member galaxies (left-hand panels) and contamination galaxies (right-hand panel) spanning median signal-to-noise $2\lesssim\mathrm{S/N/pixel}\lesssim30$. The red overplotted regions show the range of spectra encompassing the central 68\% and 95\% (dark and lighter red, respectively) of the posterior PDFs for our spectral model (\S\ref{ILSSPM}).  The text in each panel lists the median S/N and our estimates of $\vlos$, $\mathrm{Age}$, $\feh$, $\afe$, and $\disp$ as well as the ID\#'s for easy reference to the data listed in Table \ref{A267Results}. The bottom portion of each panel shows the residuals of these fits scaled by the variance in each pixel (Eq. \ref{delta residuals}).}
\label{A267BestFits}
\end{figure*}

After determining the wavelength solutions and correcting for fiber throughput and pixel sensitivity, we estimate the background sky and subtract it from our science exposures.
Apart from a strong atmospheric emission feature at $\sim5600$\AA, the main source of sky background is scattered sunlight.
Following \citet{Koposov2011} we begin by taking the sky fibers (in this set of exposures $\sim33$) for a given frame and interpolate the individual sky spectra onto a common grid with constant spacing $\Delta\lambda^\prime/\Delta\mathrm{pix}^\prime\sim0.1$\AA (oversampled by a factor of 16 with respect to the original sampling).
For each discrete wavelength of the oversampled sky spectrum, we record the median count level and estimate the variance as
\begin{equation}
\label{skyvar}
\mathrm{Var_{sky}}=2.198\pi\frac{\mathrm{MAD}^2}{2N_\mathrm{sky}}
\end{equation}
where $N_\mathrm{sky}\sim33$ and $\mathrm{MAD}$ is the median absolute deviation \citep{Rousseeuw1993}.
We then interpolate the resulting spectrum of median sky level and associated variances back onto the real, irregularly-sampled wavelength solution that is unique to a given science spectrum.
Lastly we subtract the sky spectrum from the science spectrum, pixel by pixel.

Following sky subtraction, we then combine sub-exposures by taking the inverse-variance weighted mean at each pixel using the \textsc{iraf} package, SCOMBINE with the rejection routine CRREJECT (CosmicRayREJECT) to remove cosmic rays.
Fig. \ref{A267BestFits} displays examples of the resulting M2FS spectra for science targets.

\section{Integrated Light Population Synthesis Model For Galaxy Spectra}
\label{ILSSPM}
We model the galaxy spectra by generating synthetic integrated light spectra (ILS).
This model is building on the procedure of \citet{Walker2015a} (hereafter W15b), but here we are extending this model from resolved stellar spectra to integrated light spectra.
The general procedure is to build a luminosity-weighted sum of template stellar spectra that correspond to a simple stellar population of a given age, metallicity, and alpha-abundance, and then to shift and smooth that spectrum to match the redshift, internal velocity dispersion, and instrumental broadening of the spectrograph.

\subsection{Integrated Light Spectral Library}
\label{ILS Library}

The first component of the model is a stellar spectral library.
We use the Phoenix Stellar Spectral Library \citep{Husser2013} as the basis for building the integrated light spectra.
This synthetic spectral library is computed on a regular four dimensional grid in $T_\mathrm{eff}$, $\log\ g$, [Fe/H], and [$\alpha$/Fe] space spanning a large range in each parameter: $0-15$ Gyr in Age, $0$ to $5$ in $\log\ g$, $-4.0$ to $+0.5$ in $\feh$, and $-0.2$ to $+0.8$ in $\afe$.
We continuum-normalize each spectrum beforehand.
This library does not include rare bright stars such as carbon or asymptotic giant branch (AGB) stars and therefore they will not be included in our model.
Despite their rarity, the high luminosity of these stars can contribute significantly to a galaxy's integrated light \citep{Conroy2010}.
Because our model does not include the contribution of these stars, our parameter estimates are susceptible to systematic error that is not reflected in the quoted random errors.

In order to calculate an integrated light spectrum, we need to sum know how the stars from the stellar library contribute to the light in each stellar population; in other words we need to sum luminosity-weighted contributions along the isochrone for a given stellar population.
For this we use the Dartmouth Isochrone Database \citep{Dotter2008}\footnote{http://stellar.dartmouth.edu/models/}.
This database consists of a three dimensional grid of isochrone lists in galactic age, mean metallicity [Fe/H], and chemical abundance [$\alpha$/Fe] space.
We construct a regular grid of isochrone lists with $\Delta\age=0.25\ \mathrm{Gyr}$, $\Delta\feh=0.5\ \mathrm{dex}$, and $\Delta\afe=0.2\ \mathrm{dex}$.
Each isochrone is a list of stellar properties (mass, effective temperature, magnitudes, surface gravity) describing the stars of a given age, metallicity, and chemical enrichment.

We first generate an integrated light spectrum for each isochrone, thus converting the isochrone database into an integrated light spectral library.
For this procedure, we weight each individual library spectrum according to the luminosity function computed by \citet{Dotter2008}.
For each luminosity bin in the tabulated luminosity function, we identify the isochrone having luminosity closest to the bin's central value.
For the luminosity function we use a magnitude bin width of 0.1 and a Chabrier log-normal initial mass function (IMF) of the form
\begin{equation}
\label{IMF}
dN/dM\propto \exp\left[-\frac{\ln\left(M/M_c\right)^2}{2\sigma^2}\right]
\end{equation}
where $M_c=0.22M_\odot$ is the central mass and $\sigma=0.57$ is the dispersion \citep{Chabrier2003}.
In principle these could be free parameters of our model as well, but for now we hold them fixed.
We identify which stars listed in the isochrone fall within a given magnitude bin (from the luminosity function) and determine the stellar parameters ($T_\mathrm{eff}$, $\mathrm{log\ g}$, $\feh$, and $\afe$) associated with the median star within the bin.
This star will be included in the integrated light spectrum with a weight that is simply the product of the number of stars in the magnitude bin calculated from the luminosity function and the luminosity of the star selected.

We denote the original spectra in the library as $L_0(\lambda,\bm{\theta}_\mathrm{atm})$, corresponding to stellar-atmospheric parameters $\bm{\theta}_\mathrm{atm}\equiv\left(T_\mathrm{eff},\log\ g, [\mathrm{Fe/H}], [\alpha/\mathrm{Fe}]\right)$.
As described by W15b (their equations 7 and 8), we apply a smoothing kernel over the entire stellar library to obtain a unique spectrum at the specific $\bm{\theta}_\mathrm{atm}$ of each isochrone.
We denote the smoothed spectra as $L_1(\lambda,\bm{\theta}_\mathrm{atm})$.
In our case, the number of spectra in the Phoenix library is $N_L=5566$, and we set the smoothing bandwidths equal to the grid spacing in each dimension: $h_{T_\mathrm{eff}}=200\ \mathrm{K}$, $h_\mathrm{log\ g}=0.5\ \mathrm{dex}$, $h_\mathrm{[Fe/H]}=0.5\ \mathrm{dex}$, and $h_{[\alpha/\mathrm{Fe}]}=0.2\ \mathrm{dex}$.

After generating each individual stellar spectrum $L_{1,i}\left(\lambda,\bm{\theta}_{\mathrm{atm},i}\right)$ corresponding to each isochrone, we weight and sum these spectra as described above, which produces an integrated light spectrum given by:
\begin{equation}
\label{L_ils}
L_\mathrm{ILS}\left(\lambda,\bm{\theta}_\mathrm{gal}\right)=\sum^{N_\phi}_i L_{1,i}\left(\lambda,\bm{\theta}_{\mathrm{atm},i}\right)w_i.
\end{equation}
The weight given to each spectrum is $w_i\equiv n_i\phi_i$, where  $n_i$ is the number of stars in the given magnitude bin and $\phi_i$ is the luminosity of the star specified in the isochrone; $N_\phi$ is the number of magnitude bins in the luminosity function and $\bm{\theta}_\mathrm{gal}\equiv\left(\mathrm{Age},\mathrm{[Fe/H]},[\alpha/\mathrm{Fe}]\right)$ are the galactic parameters specific to each isochrone.
We do this for the entire isochrone database, thus generating an integrated light spectral library covering the parameter space defined by $\bm{\theta_\mathrm{gal}}$.

When fitting the galactic spectra there are two processes that broaden the absorption features: instrumental line spread function (LSF) and internal motions (i.e. redshift distribution, internal velocity dispersion, galaxy rotations, and so on).
The instrumental LSF must be measured independently to break its inherent degeneracy with internal velocity dispersion (see \S\ref{Spectral Model} below).
To mimic broadening, we add another dimension to our integrated light spectral library: a smoothing parameter $h_0$.
Following the same procedure described by W15b to broaden the spectra over a range of smoothing bandwidths, we apply Eqs. 5 and 6 from W15b to each $L_\mathrm{ILS}\left(\lambda,\bm{\theta}_\mathrm{gal}\right)$.
We generate six versions of each integrated light spectrum using smoothing bandwidths $h_0=0,2,4,6,8,10$\ \AA.
This range of smoothing bandwidths was chosen so as to cover the broadening associated with the range of internal velocity dispersions we expect to measure in our galaxy sample (up to $\sim550\ \kms$ at 5500 \AA).

\subsection{Spectral Model}
\label{Spectral Model}

\begin{table*}
\centering
\caption{Free parameters and priors for Integrated Light Population Synthesis Model}
\begin{tabular}{ l l l }
\hline
\hline
Parameter &
Prior &
Description\\
\hline
$\vlos/\left(\mathrm{km\ s^{-1}}\right)$ & Uniform between $0$ and $138000$ & Line-of-sight velocity ($z=\vlos/c$)\\
$\age/\mathrm{Gyr}$ & Uniform between $0$ and $15$ & Age of simple stellar population\\
$\feh$ & Uniform between $-4$ and $+0.5$ & Metallicity of simple stellar population\\
$\afe$ & Uniform between $-0.2$ and $+0.8$ & Chemical abundance of simple stellar population\\
$\disp/\mathrm{km\ s^{-1}}$ & Uniform between $0$ and $500$ & Internal velocity dispersion of simple stellar population\\
$h_0/$\AA & Uniform between $0$ and $4$ & Polynomial coefficient (line spread function: Eq. \ref{H_n})\\ 
$h_1/$\AA & Uniform between $-2$ and $+2$ & Polynomial coefficient (line spread function: Eq.  \ref{H_n})\\ 
$h_2/$\AA & Uniform between $-4$ and $+4$ & Polynomial coefficient (line spread function: Eq.  \ref{H_n})\\ 
$p_0$ & Uniform between $-2$ and $+2$ & Polynomial coefficient (continuum: Eq. 3 from W15b)\\
$p_1$ & Uniform between $-2$ and $+2$ & Polynomial coefficient (continuum: Eq. 3 from W15b)\\
$p_2$ & Uniform between $-2$ and $+2$ & Polynomial coefficient (continuum: Eq. 3 from W15b)\\
$p_3$ & Uniform between $-2$ and $+2$ & Polynomial coefficient (continuum: Eq. 3 from W15b)\\
$p_4$ & Uniform between $-2$ and $+2$ & Polynomial coefficient (continuum: Eq. 3 from W15b)\\
$p_5$ & Uniform between $-2$ and $+2$ & Polynomial coefficient (continuum: Eq. 3 from W15b)\\
\hline
\end{tabular}
\label{A267Params}
\end{table*}

Following \citet{Koleva2009}, \citet{Koposov2011}, and W15b, we fit each individual galaxy spectrum with a spectral model of the form
\begin{equation}
\label{M_lam}
M(\lambda)=\mathrm{max}\left[S\left(\lambda\right)\right]P_l(\lambda)T\left(\lambda\left[1+\frac{v_\mathrm{los}}{c}\right]\right)
\end{equation}
where $c$ is the speed of light and $\mathrm{max}\left[S\left(\lambda\right)\right]$ is the maximum count level of a science spectrum.
Eq. \ref{M_lam} is the same as Eq. 2 in W15b, except we chose to not include the polynomial $Q_m(\lambda)$ which is a wavelength-dependent redshift.
We noticed from fitting the A267 spectra that the parameters needed for this polynomial are unconstrained in our low resolution integrated light spectra, but relatively well constrained by the high-resolution, resolved stellar spectra of W15b.
We still included in this model the same form for the polynomial $P_l(\lambda)$ given by W15b's Eqs. 3, which fits the continuum of the observed spectra.

Because we are modeling a population of stars, we build into our model a way of measuring the internal velocity dispersion of this population.
The velocity dispersion will manifest itself as a broadening of the absorption features in each spectrum.
However, this broadening will be degenerate with the line spread function (LSF) of the spectrograph, so care must be taken to break this degeneracy between the two sources of broadening.
To do this, we first measure the LSF with twilight spectra and then broaden the model spectra according to the LSF in addition to the broadening associated with the velocity dispersion of the stars.
In order to allow for a wavelength dependent LSF, we introduce another polynomial for the smoothing bandwidth, $H_n(\lambda)$, which we allow to vary with wavelength according to
\begin{align}
\label{H_n}
H_n(\lambda)=h_0+h_1\left[\frac{\lambda-\lambda_0}{\lambda_s}\right]+h_2\left[\frac{\lambda-\lambda_0}{\lambda_s}\right]^2&\nonumber\\+...+h_n\left[\frac{\lambda-\lambda_0}{\lambda_s}\right]^n&.
\end{align}
Given that the broadening related to velocity dispersion $\sigma_\mathrm{int}$ is given by $\sigma_\mathrm{int}/c=\Delta\lambda/\lambda$ where $c$ is the speed of light, the total broadening associated with both the LSF and the internal velocity dispersion of the population of stars is given by
\begin{equation}
\label{h_lam}
h^2(\lambda)=\left(\frac{\sigma_\mathrm{int}}{c}\lambda\right)^2+H_n^2(\lambda).
\end{equation}
This method introduces $n+2$ new free parameters: one for internal velocity dispersion and the other $n+1$ are from the $h_n$ coefficients.
However, when fitting twilight spectra, we assume that $\disp=0$ because the ``stellar population" consists of only the sun, and so we only fit the $n+1$ parameters associated with the LSF.
On the other hand, when fitting science spectra we use the previously measured $n+1$ LSF parameters from the twilight fits, and so we fit only for $\disp$.

In order to let the spectral model vary continuously despite the library's coarse gridding in galactic parameter space and the discrete values of the smoothing bandwidth, we apply another wavelength-dependent smoothing over the entire collection of library spectra.
Specifically, for any choice of galactic parameters $\bm{\theta}_\mathrm{gal}$ and smoothing bandwidth $h(\lambda)$, we obtain a unique template
\begin{equation}
\label{T_lam}
T(\lambda)=\frac{\sum\limits_i^{N_\mathrm{ILS}}L_\mathrm{ILS}\left(\lambda,\bm{\theta}_{\mathrm{gal}_i},h_{0_i}\right) K_3\left(\lambda,\frac{\bm{\theta}_{\mathrm{gal}_i}-\bm{\theta}_\mathrm{gal}}{\bm{h}_\mathrm{gal}},\frac{h_{0_i}-h(\lambda)}{h_h}\right)}{\sum\limits^{N_\lambda}_iK_3\left(\lambda,\frac{\bm{\theta}_{\mathrm{gal}_i}-\bm{\theta}_\mathrm{gal}}{\bm{h}_\mathrm{gal}},\frac{h_{0_i}-h(\lambda)}{h_h}\right)}
\end{equation}
where $N_\mathrm{ILS}=14202$ is the number of ILS library spectra and the kernel is
\begin{multline}
\label{K3}
K_3\left(\lambda,\frac{\bm{\theta}_{\mathrm{gal}_i}-\bm{\theta}_\mathrm{gal}}{\bm{h}_\mathrm{gal}},\frac{h_{0_i}-h(\lambda)}{h_h}\right)=\\
\exp\left[-\frac12
\left(\frac{\left(\mathrm{Age}_i-\mathrm{Age}\right)^2}{h_\mathrm{Age}^2}+
\frac{\left(\mathrm{[Fe/H]}_i-\mathrm{[Fe/H]}\right)^2}{h_\mathrm{[Fe/H]}^2}+\right.\right.\\
\left.{}\left.{}\frac{\left([\alpha/\mathrm{Fe}]_i-[\alpha/\mathrm{Fe}]\right)^2}{h_{\afe}^2}+
\frac{\left(h_{0_i}-h(\lambda)\right)^2}{h_h^2}\right)
\right].
\end{multline}
We set the galactic smoothing bandwidths $\bm{h}_\mathrm{gal}$ equal to the grid spacing in each dimension: $h_\mathrm{Age}=0.25\ \mathrm{GYr}$, $h_\feh=0.5\ \dex$, $h_{\afe}=0.2\ \dex$, $h_h=1$\AA.
We found that setting the smoothing bandwidth $h_h=2$\AA\ (the grid spacing of the library) results in our model favoring a larger broadening parameter, thus over-smoothing the spectral fit; therefore, we decreased the smoothing bandwidth to $h_h=1$\AA, which solved this issue.
This smoothing procedure gives posterior probability distributions that are approximately Gaussian and tend not to cluster near the library's grid points.

\section{Analysis of Spectra}
\label{Analysis of Spectra}

We now apply this model for fitting spectra and estimating model parameters.

\subsection{Likelihood function and free parameters}
\label{lhood}

Given the spectral model $M(\lambda)$, we assume that the observed spectrum $S(\lambda)$ has likelihood
\begin{multline}
\label{lhood func}
\mathcal{L}\left(S(\lambda)|\bm{\theta}\right)=\\
\prod^{N_\lambda}_{i=1}\frac{1}{\sqrt{2\pi\mathrm{Var}\left[S(\lambda_i)\right]}}\exp{\left[-\frac12\frac{\left(S(\lambda_i)-M(\lambda_i)\right)^2}{\mathrm{Var}\left[S(\lambda_i)\right]}\right]}.
\end{multline}
In practice the value of $M(\lambda_i)$ that we use in Eq. \ref{lhood func} is the linear interpolation, at observed wavelength $\lambda_i$, of the discrete model we calculate from Eq. \ref{M_lam}.
This interpolation is necessary because a given template spectrum $T(\lambda)$ retains the discrete wavelength sampling of the synthetic library, which generally differs from those of the observed spectra.

Following W15b in order to define the polynomials in Eqs. \ref{M_lam} and \ref{H_n}, we chose $l=5$ and $n=2$, respectively.
These choices give sufficient flexibility to fit the continuum shape and to apply low-order corrections to the wavelength solution.
We adopt scale parameters $\lambda_0$ and $\lambda_s$ such that $-1\le(\lambda-\lambda_0)/\lambda_s\le1$ over the entire wavelength range considered in a fit.

With these choices the spectral model $M(\lambda)$ is fully specified by a vector of 14 free parameters:
\begin{multline}
\label{theta}
\bm{\theta}=(\vlos,\ \age,\ \feh,\ \afe,\ \disp,\ h_0,\ h_1,\ h_2,\\
p_0,\ p_1,\ p_2,\ p_3,\ p_4,\ p_5).
\end{multline}
The first five have physical meaning and the rest are nuisance parameters.
Table \ref{A267Params} lists all parameters and identifies the adopted priors, all of which are uniform over the specified range of values and zero outside that range.

\subsection{Parameter Estimation}
\label{param est}

From Bayes' theorem, given the observed spectrum $S(\lambda)$, the model has a posterior probability distribution function (PDF)
\begin{equation}
\label{bayes}
p\Big(\bm{\theta}|S(\lambda)\Big)=\frac{\mathcal{L}\Big(S(\lambda)|\bm{\theta}\Big)p(\bm{\theta})}{p\Big(S(\lambda)\Big)}
\end{equation}
where $\mathcal{L}\Big(S(\lambda)|,\bm{\theta}\Big)$ is the likelihood from Eq. \ref{lhood func}, $p(\bm{\theta})$ is the prior and
\begin{equation}
\label{evidence}
p\Big(S(\lambda)\Big)=\int\mathcal{L}\Big(S(\lambda)|\bm{\theta}\Big)p(\bm{\theta})\mathrm{d}\bm{\theta}\mathrm{d}s_1\mathrm{d}s_2
\end{equation}
is the marginal likelihood, or `evidence'.

In order to evaluate the posterior PDF, we must scan the 14D parameter space.
For this task, we use the software package \textsc{multinest}\footnote{Available at ccpforge.cse.rl.ac.uk/gf/project/multinest} \citep{Feroz2008, Feroz2009}.
\textsc{multinest} implements a nested-sampling Monte Carlo algorithm that is designed to calculate the evidence (Eq. \ref{evidence}) and simultaneously to sample the posterior PDF (Eq. \ref{bayes}).
\citet{Feroz2008} and \citet{Feroz2009} demonstrate that \textsc{multinest} performs well even when the posterior is multimodal and has strong curving degeneracies, circumstances that can present problems for standard Markov Chain Monte Carlo techniques.

\subsection{Tests with Mock Spectra}
\label{mock spectra}

\begin{table}
\centering
\caption{Input physical parameters for the mock spectral catalog.}
\label{MockParams}
\begin{tabular}{ c c c c c }
\hline
\hline
$\vlos$ & $\age$ & $\feh$ & $\afe$ & $\disp$\\
km s$^{-1}$  & Gyr & dex & dex & km s$^{-1}$\\
 \hline
 69579 & 10.9 & 0.39 & -0.03 & 540\\
74406 & 5.8 & -0.52 & 0.68 & 472 \\
71331 & 9.0 & -2.11 & 0.79 & 194\\
70380 & 10.8 & -0.12 & 0.24 & 149\\
73100 & 13.2 & -0.39 & 0.06 & 367\\
73851 & 14.9 & -1.38 & 0.02 & 271\\
67592 & 0.2 & -0.64 & 0.32 & 466\\
73442 & 2.8 & -0.92 & 0.27 & 256\\
67467 & 3.1 & -1.86 & 0.13 & 489\\
72545 & 7.9 & -0.42 & 0.17 & 296\\
69134 & 8.7 & 0.60 & -0.15 & 535\\
69698 & 1.4 & -0.36 & -0.10 & 204\\
68629 & 7.9 & 0.36 & 0.08 & 287\\
68724 & 14.4 & 0.36 & 0.20 & 519\\
71461 & 12.5 & -2.38 & 0.63 & 507\\
68701 & 13.5 & -0.38 & 0.58 & 534\\
73772 & 13.8 & 0.35 & 0.09 & 319\\
67092 & 1.4 & -1.60 & 0.68 & 144\\
65444 & 3.2 & -2.13 & -0.04 & 365\\
67506 & 12.0 & -2.13 & 0.51 & 364\\
\hline
\end{tabular}
\end{table}

\begin{figure*}
\centering
\includegraphics[width=.49\textwidth]{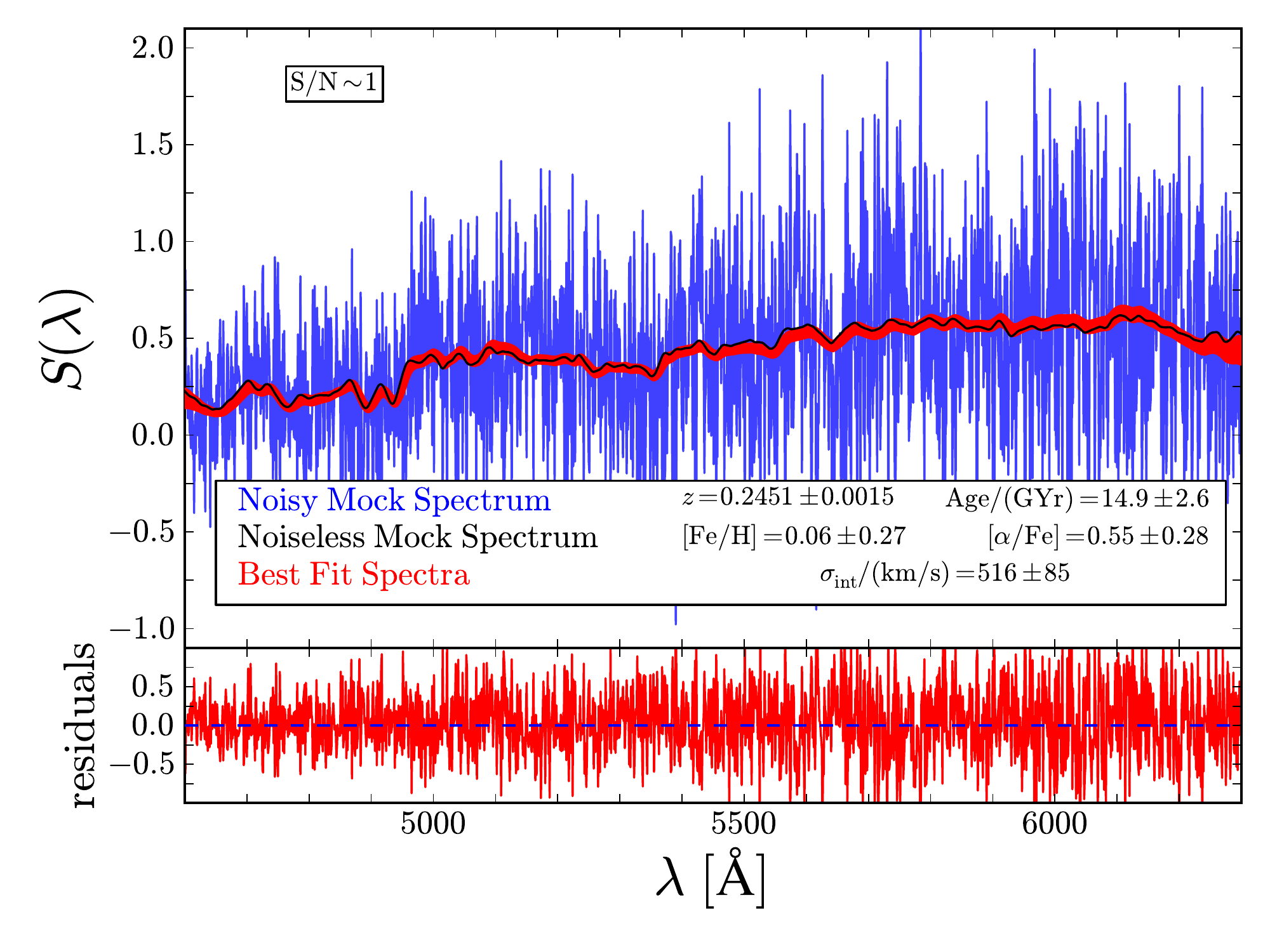}
\includegraphics[width=.49\textwidth]{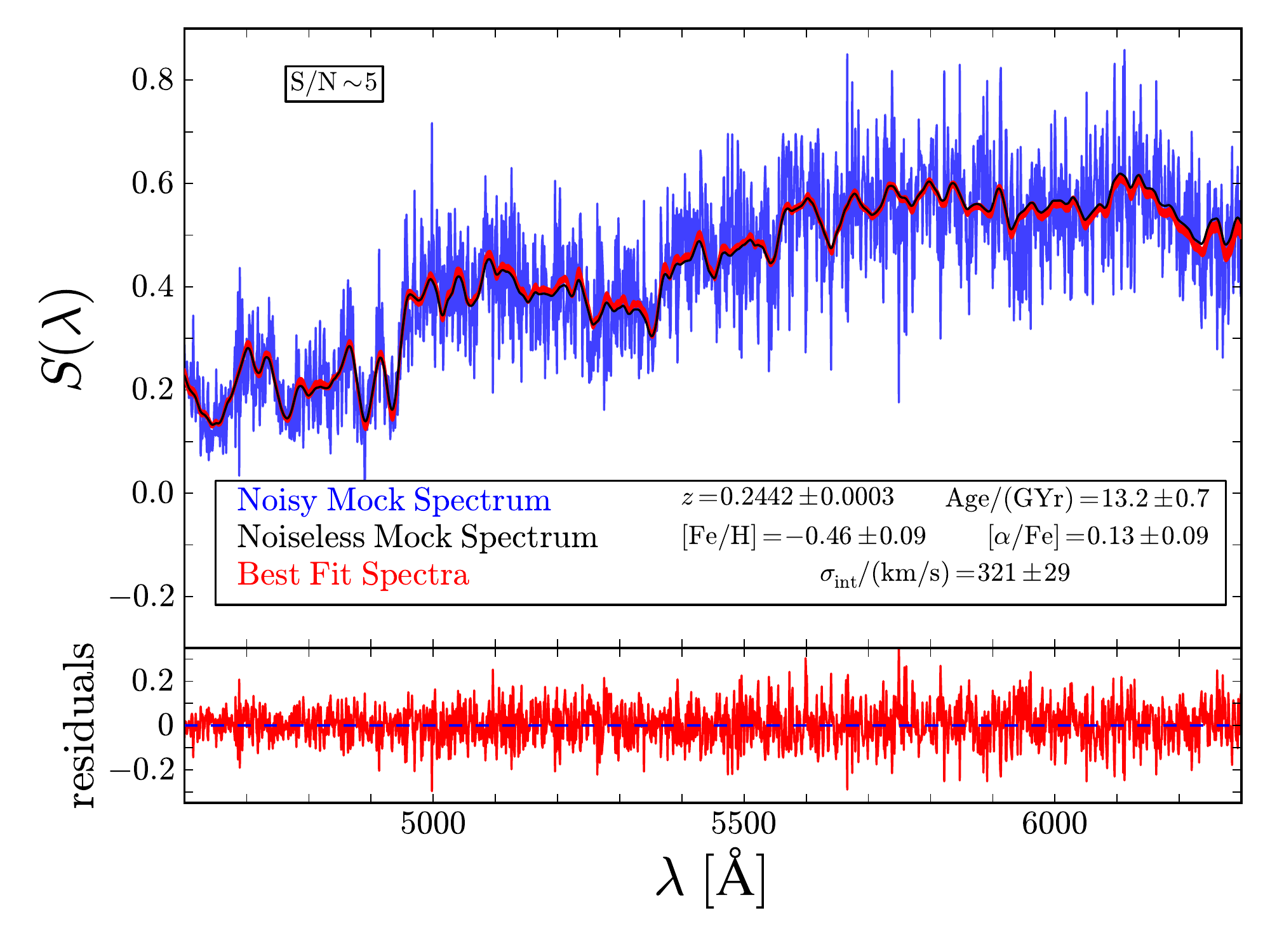}
\includegraphics[width=.49\textwidth]{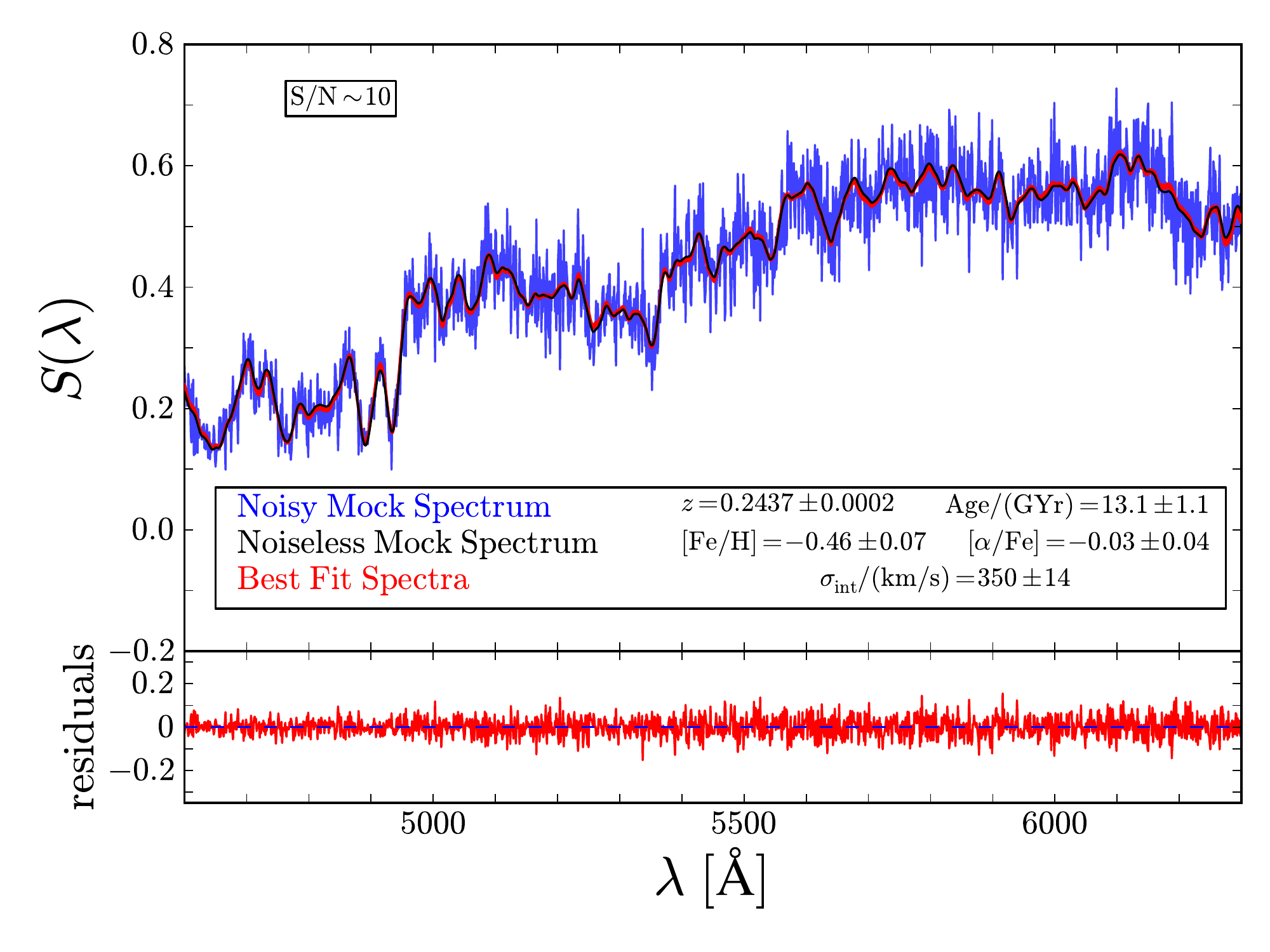}
\includegraphics[width=.49\textwidth]{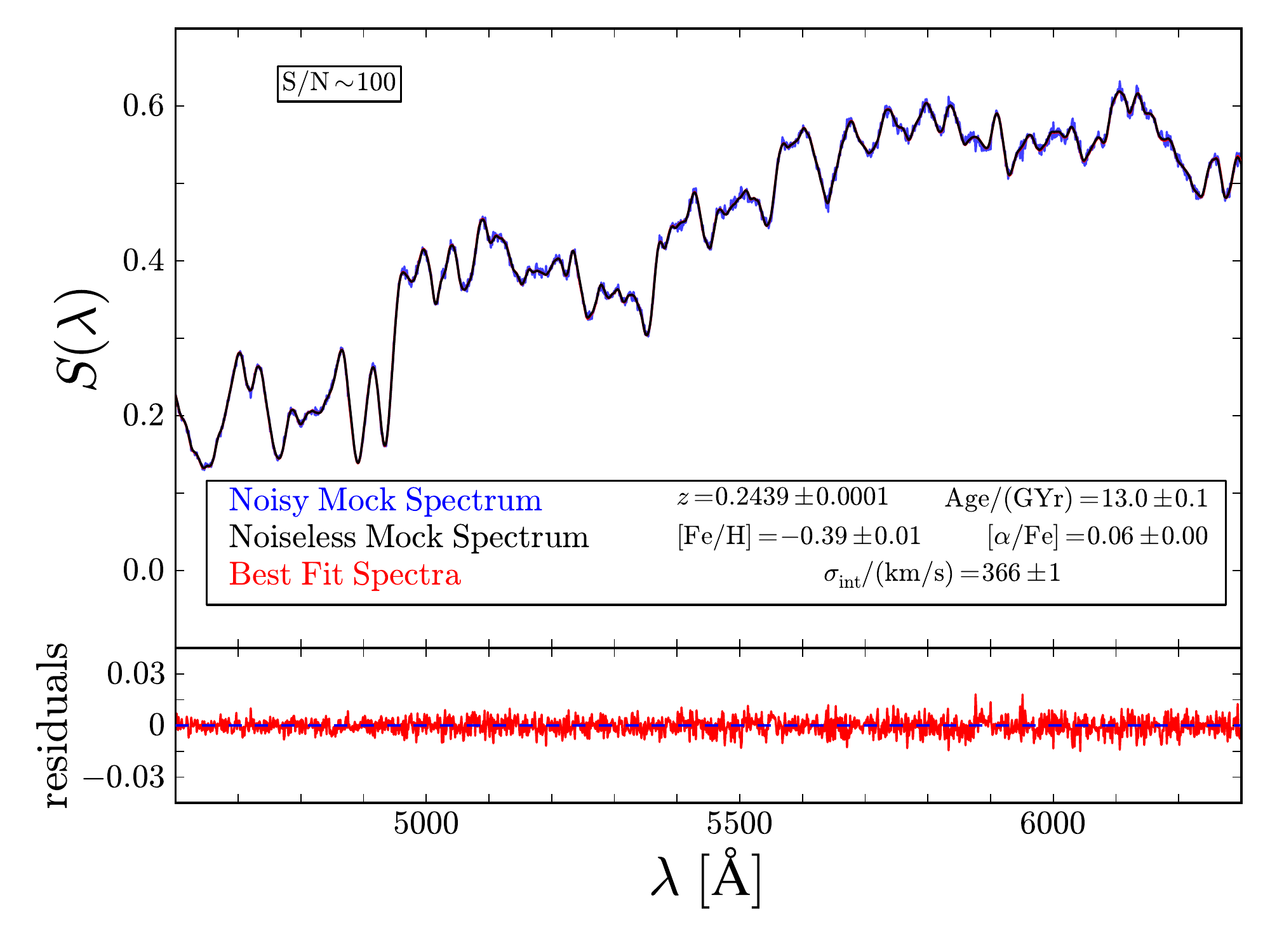}
\caption{Mock spectra (blue) for different values of median S/N per pixel, which is identified in the top left of each panel.  The original, noiseless mock spectrum is plotted in each panel in black.  Over plotted in red in the top portion of each panel is the range of spectra encompassing 68\% of the posterior distribution of the spectral fit.  Also in each panel we list the best fit parameters with uncertainties.  For $\vlos$ we show redshift $z$ instead.  In the bottom portion of each panel, we show the residual difference between the noisy mock spectrum and the best fit spectrum from the model.}
\label{MockBestFits}
\end{figure*}

\begin{figure}
\centering
\includegraphics[width=\columnwidth]{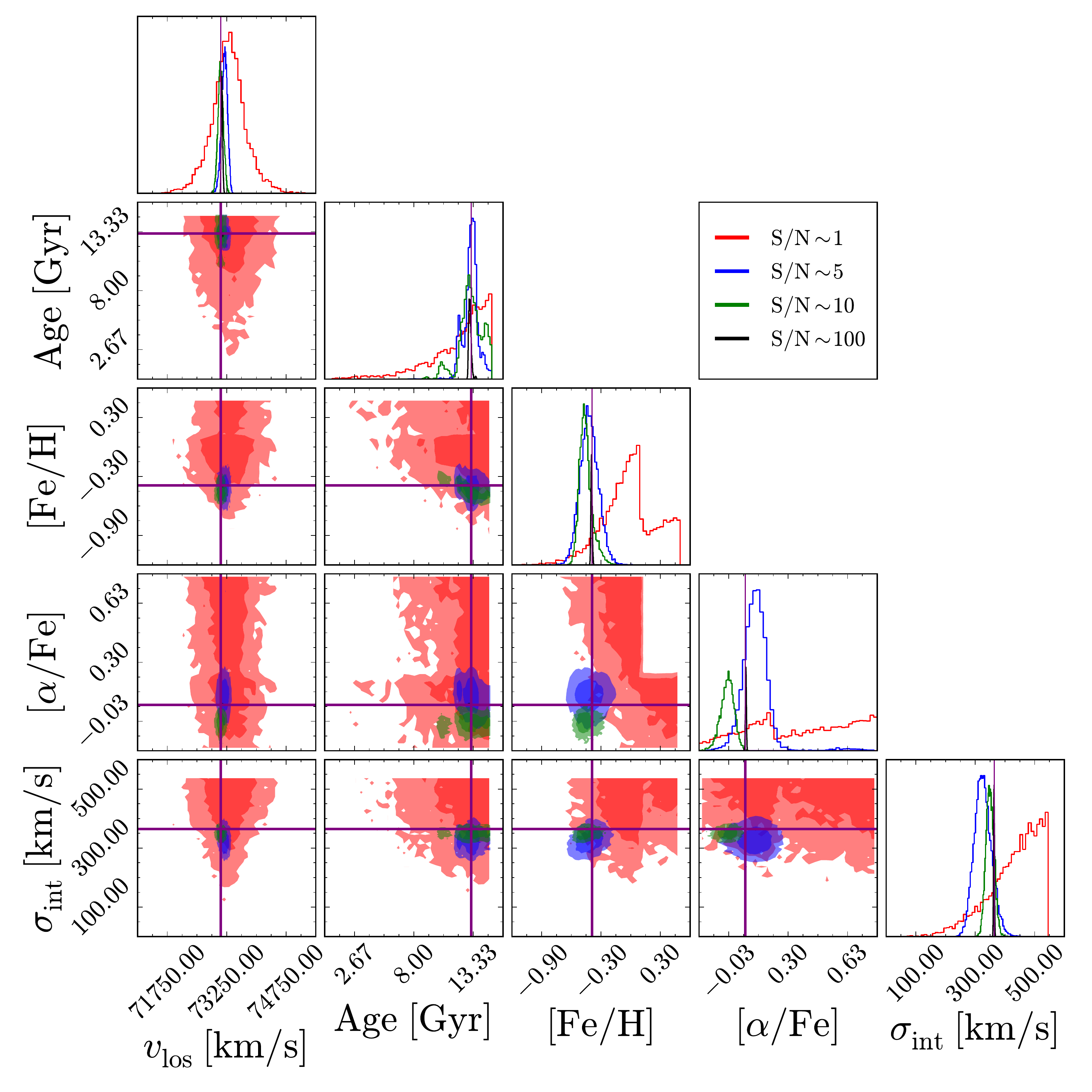}
\caption{Marginal posterior probability distributions for the five galactic parameters corresponding to the fits to mock spectra shown in Fig. \ref{MockBestFits}.  Each S/N value is represented with a different color as indicated in the top right panel.  For the 2D posteriors, we show the the 1$\sigma$ and 2$\sigma$ regions of these distributions as the darker and lighter regions respectively. Above each column is the marginalized 1D posterior PDFs for each of the five parameters. Also shown in each panel in purple is the input value of the parameters used in generating this noiseless mock spectrum.}
\label{MockPosteriors}
\end{figure}

\begin{figure}
\centering
\includegraphics[width=\columnwidth]{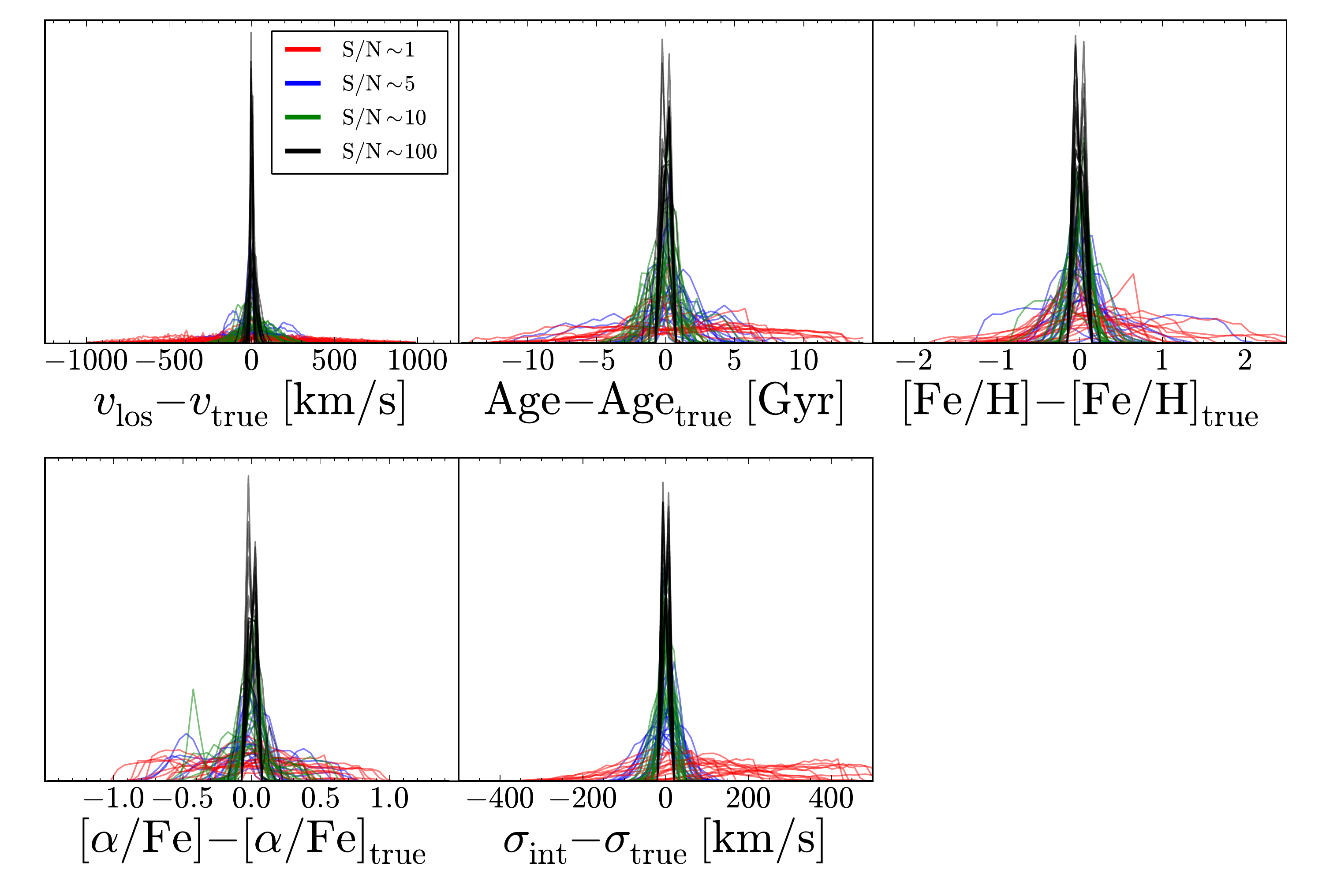}
\caption{Difference between all posterior PDFs and the true input values (subscript true) for the five physical parameters for all mock spectra. Each PDF is colored by the median pixel S/N as shown in the top left panel.}
\label{MockSummary}
\end{figure}

As a first test of the accuracy of our model, we generated and fit mock spectra over a range of S/N values.  
We first generated a noiseless mock spectrum, for a given set of $\age$, $\feh$, $\afe$, and $\disp$, using the pre-calculated spectral library (see \S \ref{ILS Library}) and the spectral model described in \S \ref{Spectral Model}.
Table \ref{MockParams} shows the 20 sets of input galactic parameters we used to generate each noiseless mock spectrum. 
In order to also analyze the performance of our model at different S/N levels, we added noise such that the median S/N of each mock spectrum had values $\sim 1, 5, 10, 100$.
Therefore, each noiseless mock spectrum produced four noisy spectra which we fit with our model.
Each spectrum shown in Fig. \ref{MockBestFits} was generated from the same noiseless mock spectrum (the input parameters for the mock spectra shown in Fig. \ref{MockBestFits} are given in the first row in Table \ref{MockParams}).

Plotted over each mock spectrum in Fig. \ref{MockBestFits} is the best-fitting model spectrum.
We show in red the range of spectra attributed to central 68\% of the posterior distribution estimated by \textsc{multinest} at each pixel.
For high S/N levels, these red regions look just like single curves because the fits are tightly constrained; however, for low S/N, one can see the width of these distributions (top left panel of Fig. \ref{MockBestFits}).
In the bottom portion of each panel, we show the residual difference between the best fitting spectrum (most likely set of parameters) and the mock spectrum.
The text within each panel indicates estimates of physical parameters redshift $z$, $\age$, $\feh$, $\afe$, and internal velocity dispersion $\disp$.

In the text of Fig. \ref{MockBestFits}, along with the best fit values of the galactic parameters, we also list their respective uncertainties.
These uncertainties enclose the central 68\% of the posterior PDF for each parameter.
Therefore, in Fig. \ref{MockPosteriors} we show the marginal posterior PDFs for the five galactic parameters, which better quantifies the distribution of each parameter.
The posteriors shown in Fig. \ref{MockPosteriors} correspond to the spectral fits shown in Fig. \ref{MockBestFits}.
Each color in the 1D and 2D posteriors corresponds to a different median pixel S/N.
The darker and lighter regions in the 2D posteriors show the $1\sigma$ and $2\sigma$ contours of these distributions, respectively.
For increasing S/N, the posterior distributions become more Gaussian in shape (which is expected considering we use a Gaussian likelihood function Eq. \ref{lhood func}) and the 2D posteriors are much better constrained.
Furthermore, we also show the true input values of each of these parameters in purple.
We can easily see how the posteriors converge on the true values as S/N increases.
Additionally we can see that some parameters are better constrained about the true values at lower S/N ($\vlos$ for example) while other parameters ($\afe$) have difficulty at low S/N.

We repeated this test for 20 sets of input parameters (Table \ref{MockParams}), and thus a total of 80 mock spectra.
Fig. \ref{MockSummary} shows a summary of our results for the mock catalog.
In each panel of Fig. \ref{MockSummary}, we show the difference between the true input value for each mock spectrum and the posterior PDFs for each of the five galactic parameters.
Each panel corresponds to a different parameter and the colors show how these distributions vary with S/N.
As expected, with increasing S/N, these posteriors become more constrained and are more centered on zero deviation (in other words centered on the true input value for the mock spectrum).

These tests establish good statistical properties for our model.
However, they leave our estimates susceptible to systematic errors due to the choice of spectral library (e.g incomplete line list) and isochrone databases (e.g. IMF).
W15b found that there is a significant dependence on choice of spectral library, such that estimates of $\feh$ and $\afe$ can suffer systematic errors of up to $\sim0.5\dex$.
In order to gauge the magnitude of these errors, we compare results obtained from our procedure to those obtained by others using different methods.

\subsection{External Tests}
\label{External Tests}

\begin{figure}
\centering
\includegraphics[width=.49\columnwidth]{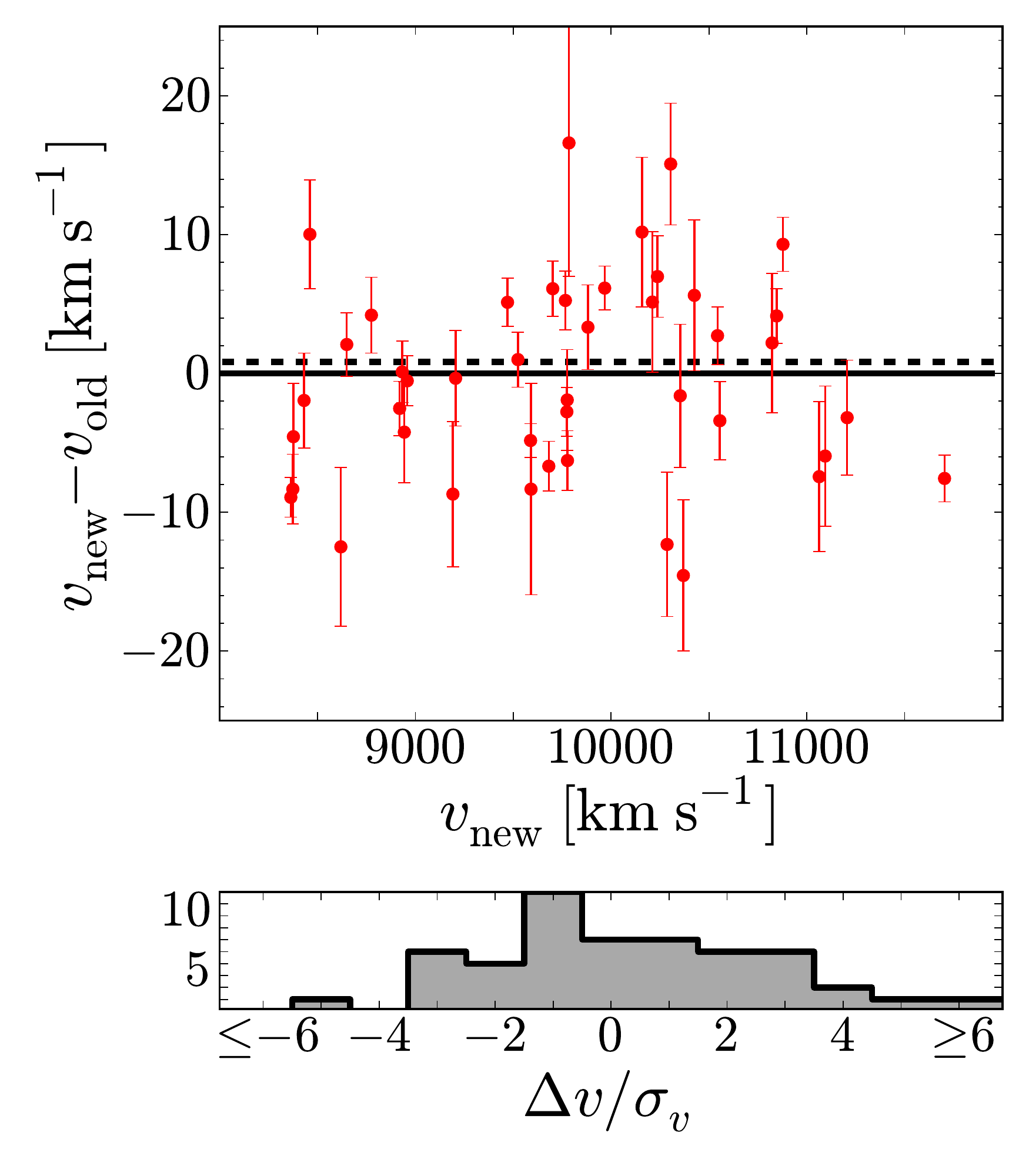}
\includegraphics[width=.49\columnwidth]{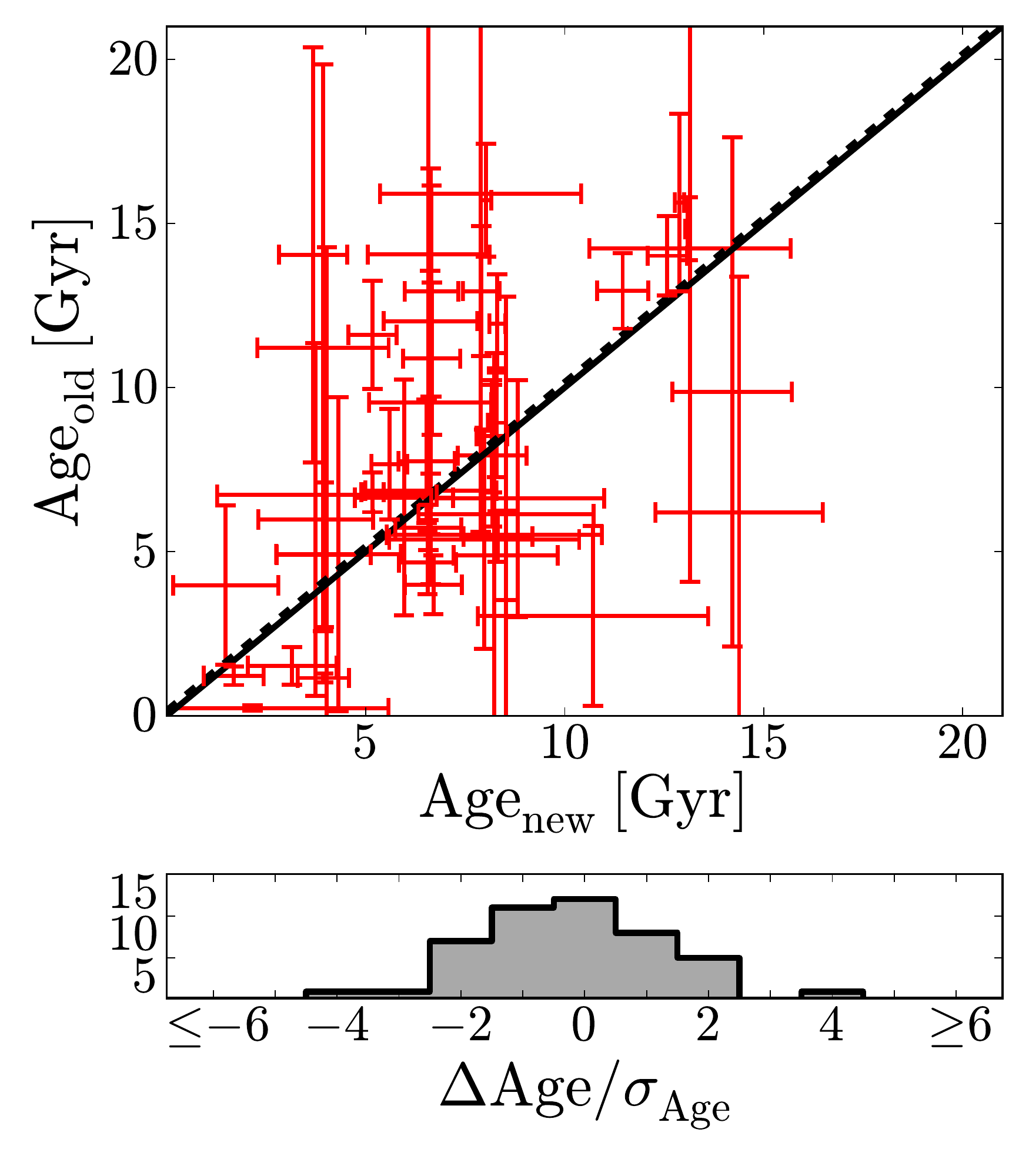}\\
\includegraphics[width=.49\columnwidth]{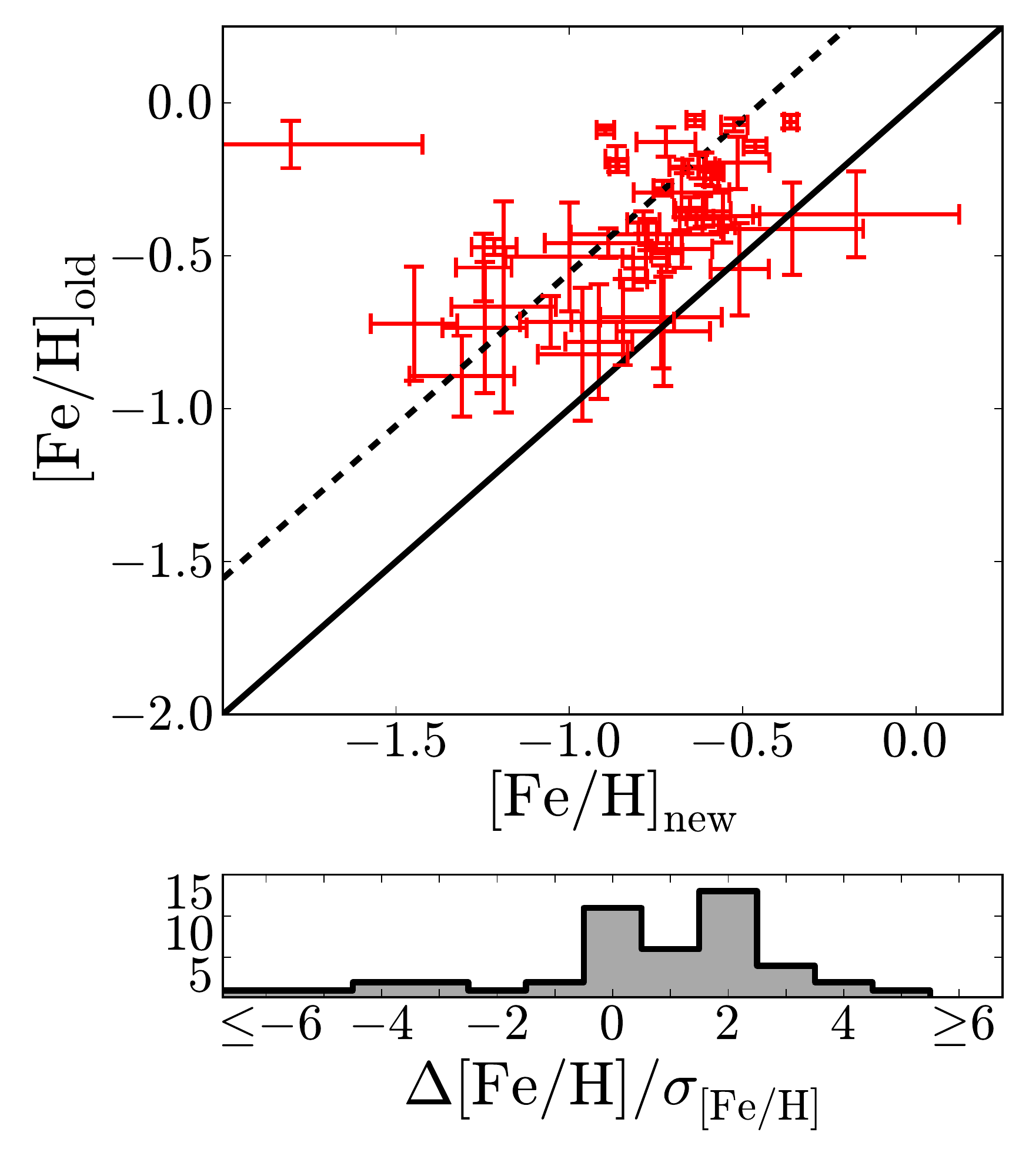}
\includegraphics[width=.49\columnwidth]{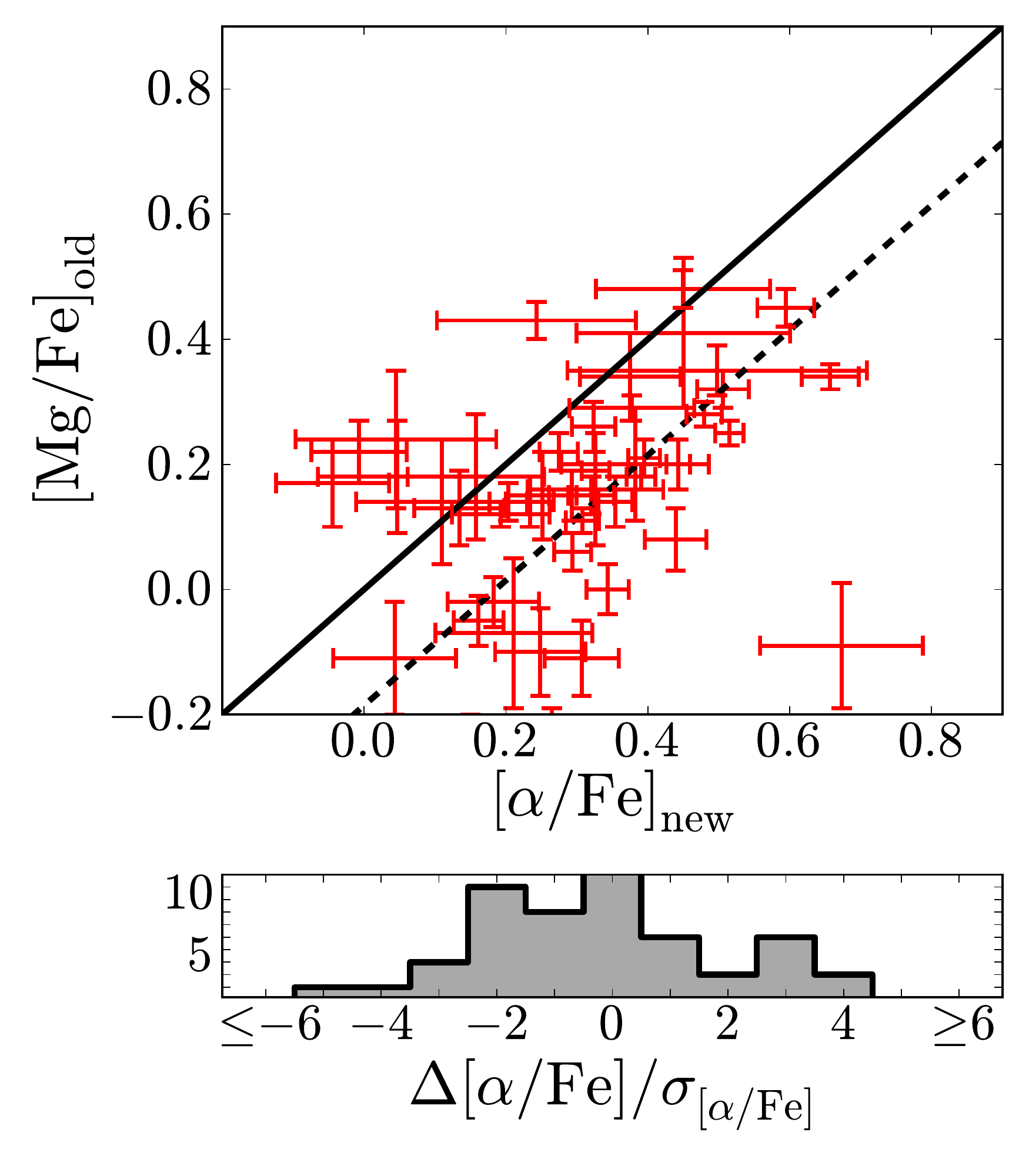}
\caption{We compare our results (subscript ``new") to those cited in \citet{Chilingarian2008} (subscript ``old") for the Abell 496 cluster.  Each group of plots corresponds to one of the galactic parameters.  The solid black line guides the eye to a one-to-one correlation between the two sets of results.  The dashed black line is a fit to the correlation keeping the slope set to one but allowing a constant systematic offset between the two sets. For the velocity panel we show the difference between the measurements in order to more clearly show their uncertainties. The histograms in the bottom panels of each plot show the difference between our measured value and the previous value after applying this constant systematic offset and scaling by the total variance in the two results.}
\label{A496Compare}
\end{figure}

As a final test of our model,  we compared our model estimates with previously published results.
The spectra for this test were generously provided by I. Chilingarian (private communications) and we compared our results to those in \citet{Chilingarian2008} (hereafter C08).
These spectra were observed on the ESO Very Large Telescope using the FLAMES/Giraffe instrument at a resolution of $R\sim6300$ in the wavelength range $5010-5831$\AA.
Following the method outlined in \citet{Chilingarian2007}, their spectral fitting method is built upon the PEGASE.HR synthetic spectra \citep{LeBorgne2004}.
Using a Salpeter IMF, they generated a template spectrum from a linear combination of synthetic spectra at a given age and metallicity similar to our procedure.
Using a multidimensional $\chi^2$ minimization procedure, they first fit the kinematics and continuum for each spectrum at a set of fixed values for age and metallicity.
Finally they obtained a map of minimal $\chi^2$ in age-metallicity space for each spectrum, from which they estimate age and metallicity for the given stellar population.
Therefore, they estimated the stellar population parameters of age and mean metallicity along with line-of-sight velocity and internal velocity dispersion, all of which we compare to the output from our model.
Furthermore, they measured Lick indices to compute magnesium abundance ratios $\mathrm{[Mg/Fe]}$ which we compare to our estimates of chemical enrichment $\afe$.

Before fitting the spectra, we noticed from manual inspection that one spectrum had strong emission lines, which we masked by setting the variance in those pixels to large values ($10^9$).
Fig. \ref{A496Compare} compares the results between the two models: our results are on the x-axis (with subscripts \textit{new}) while C08 results are on the y-axis (with subscripts \textit{old}).
The solid black line over plotted in each panel guides the eye to a one-to-one relationship.
For the velocity panel, we show the difference between measured velocities in order to more clearly show the distribution.
We also fit a linear least squares line to these distributions while fixing the slope to unity so that we can quantify any systematic differences between the two models.
These fits are shown as the dashed black lines in each panel.
In the bottom plot of each panel we, show histograms of the differences between the two models, incorporating this systematic offset, and scaled by the total uncertainty in the measurements.
In the bottom left panel we compare our measurements of chemical abundance $\afe$ to their measurement of [Mg/Fe]; therefore, this systematic offset partially correlates to the abundance of elements other than magnesium in the stellar population.
The systematic offsets between our results and theirs is most likely due to differences in the choice of spectral libraries.
We discussed above in \S\ref{mock spectra} that different library spectra can affect the stellar property estimates by up to 0.5 dex.
We caution the reader to understand that our estimates are susceptible to such systematic offsets.

The histograms show that our measured model parameters are mostly within $\sim2$ standard deviations of those measured in C08.
There are a few outliers (one most notable in $\feh$ space) which differ by $\gtrsim3\sigma$ from the values cited in C08 after accounting for systematic offsets.
These outliers are fits to low S/N spectra, and our model still produces good fits to the data even though our best fitting parameters differ from C08.
Nevertheless, it is not surprising to see one or two $3\sigma$ outliers in a sample of $\sim50$.
The distribution of the age comparisons appears to show little correlation; however, the histogram in that panel shows that our results are consistent with C08 given the cited uncertainties.
We would like to note that lacking the twilight spectra that would be necessary to estimate the instrumental LSF of C08, we do not compare $\disp$ for their spectra.

\section{Results for Abell 267}
\label{results}

\begin{table*}
\caption{Results for fitting of A267 spectra. Full table will be available online.}
\label{A267Results}
\begin{tabular}{ c c c c c c c c c c c }
\hline
\hline\\
ID & $\alpha_{2000}$ & $\delta_{2000}$ & r & i & S/N & $\overline{\vlos}$ & $\overline{\age}$ & $\overline{\feh}$ & $\overline{\afe}$ & $\overline{\disp}$\\
& (hh:mm:ss) & ($\degr$:$\arcmin$:$\arcsec$) & [mag] & [mag] & & (km s$^{-1}$) & (Gyr) & (dex) & (dex) & (km s$^{-1}$)\\\\
\hline
1 & 01:53:13.48 & +01:00:48.6 & $20.38$ & $19.86$ & $10.1$ & $114156\pm16$ & $5.5\pm0.6$ & $-0.72\pm0.10$ & $0.05\pm0.05$ & $193.3\pm12.2$\\
2 & 01:53:20.26 & +01:01:17.0 & $20.52$ & $20.06$ & $9.3$ & $114059\pm18$ & $6.5\pm0.4$ & $-2.54\pm0.05$ & $0.48\pm0.09$ & $71.6\pm24.5$\\
7 & 01:53:18.66 & +01:05:8.6 & $20.12$ & $19.64$ & $15.5$ & $83304\pm14$ & $13.5\pm1.0$ & $-1.32\pm0.07$ & $0.34\pm0.04$ & $165.1\pm13.0$\\
11 & 01:53:28.58 & +01:01:57.6 & $19.27$ & $18.79$ & $30.7$ & $54872\pm7$ & $6.5\pm0.3$ & $-0.80\pm0.04$ & $0.31\pm0.03$ & $160.3\pm6.1$\\
13 & 01:53:36.88 & +01:03:50.8 & $20.13$ & $19.63$ & $5.0$ & $66669\pm104$ & $11.8\pm2.4$ & $-1.29\pm0.25$ & $0.34\pm0.18$ & $298.6\pm84.3$\\
19 & 01:52:59.68 & +01:14:9.4 & $19.79$ & $19.27$ & $19.8$ & $69207\pm11$ & $11.4\pm0.8$ & $-1.13\pm0.06$ & $0.20\pm0.04$ & $151.8\pm10.0$\\
46 & 01:52:48.44 & +00:58:44.8 & $19.43$ & $18.91$ & $31.2$ & $69090\pm8$ & $11.7\pm0.5$ & $-1.19\pm0.04$ & $0.16\pm0.02$ & $218.2\pm9.4$\\
55 & 01:52:31.17 & +01:00:6.2 & $20.42$ & $19.84$ & $2.2$ & $66649\pm52$ & $11.4\pm2.6$ & $-0.59\pm0.25$ & $0.13\pm0.19$ & $110.5\pm52.6$\\
60 & 01:52:37.42 & +00:59:2.2 & $19.62$ & $19.06$ & $15.8$ & $66367\pm14$ & $10.5\pm1.0$ & $-1.53\pm0.08$ & $0.47\pm0.05$ & $132.0\pm12.5$\\
75 & 01:52:20.13 & +00:54:18.7 & $20.73$ & $20.26$ & $3.9$ & $118450\pm41$ & $4.4\pm0.8$ & $-2.04\pm0.27$ & $0.67\pm0.10$ & $133.8\pm33.3$\\
\hline\hline\end{tabular}\end{table*}

\begin{figure}
\centering
\includegraphics[width=\columnwidth]{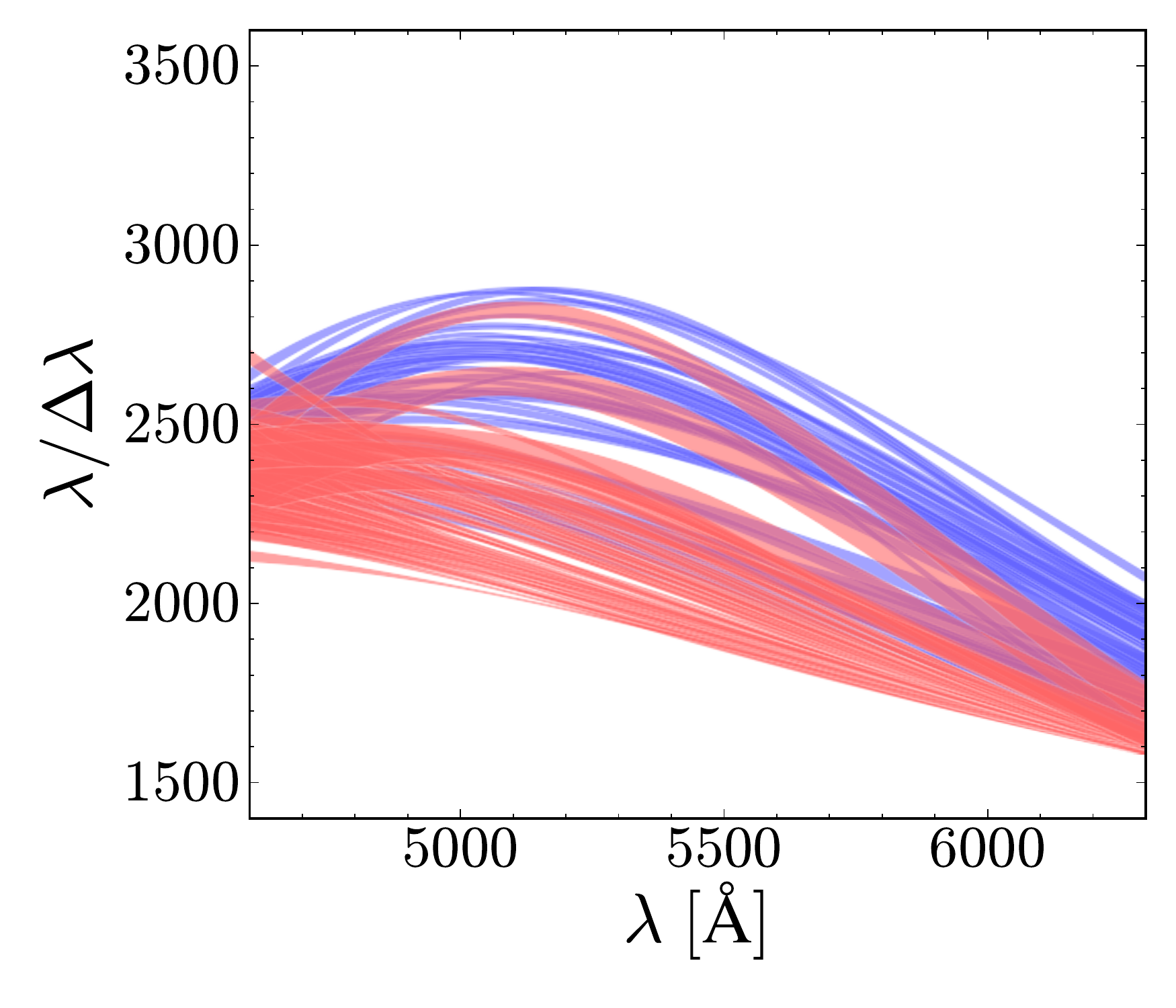}
\label{AllLSFs}
\caption{All line-spread functions measured from fitting the twilight spectra.  Each curve corresponds to the LSF measured for that given fiber.  The two colors differentiate between the two spectrographs that the fibers feed into on M2FS.  Instead of plotting a single curve for each fiber's LSF, we show the 68\% spread of each LSF as predicted by their respective posterior PDF.}
\end{figure}

\begin{figure}
\centering
\includegraphics[width=\columnwidth]{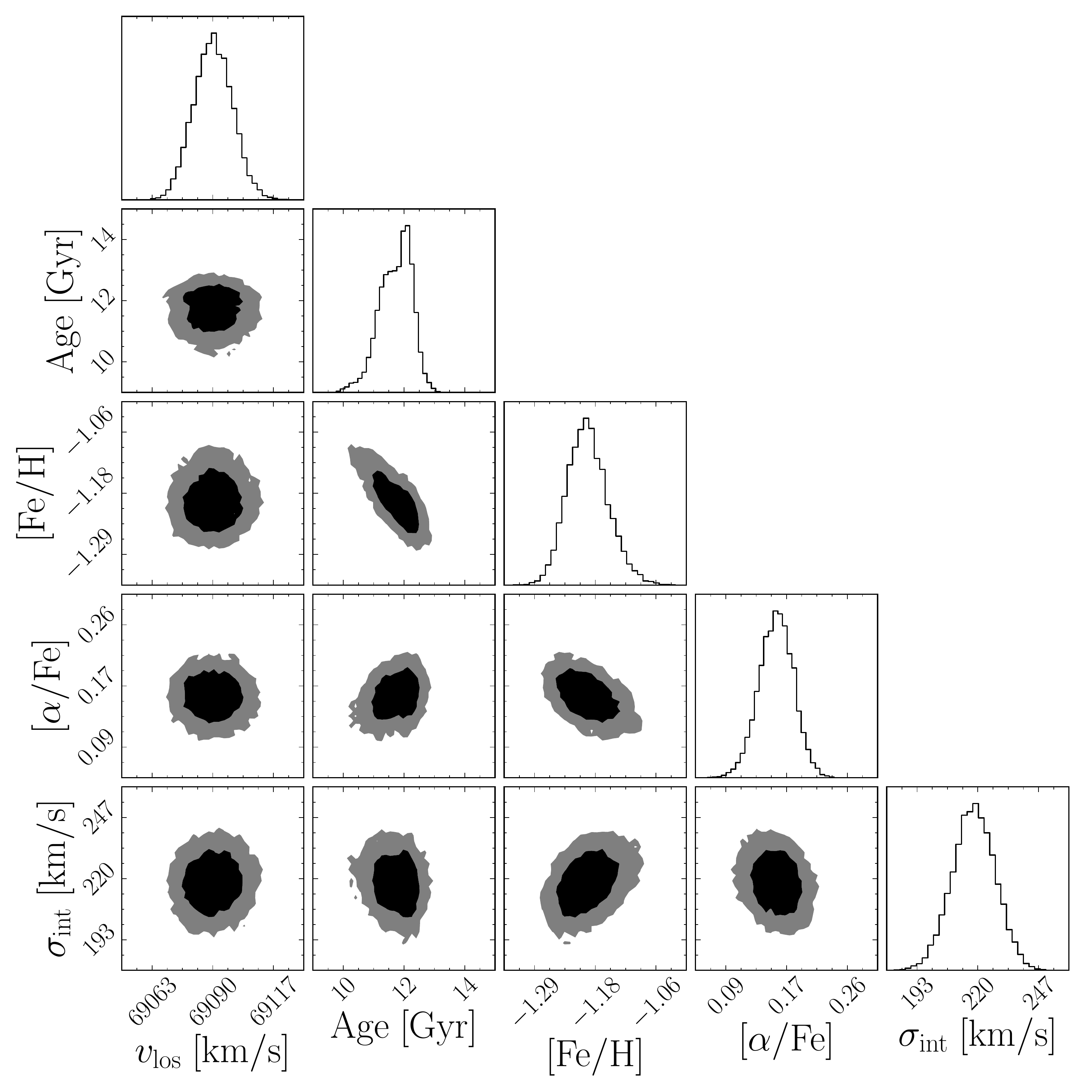}
\caption{1D and 2D posterior probability distribution functions for the five galactic parameters estimated for one of our A267 science targets (ID\#46 in Table \ref{A267Results}): line-of-sight velocity $\vlos$, age, metallicity $\feh$, chemical abundance $\afe$, and internal velocity dispersion $\disp$ of the simple stellar population.  The dark and lighter shaded regions show the $1\sigma$ and $2\sigma$ widths of the 2D marginal posterior PDFs, respectively.}
\label{A267PosteriorPart}
\end{figure}

\begin{figure}
\centering
\includegraphics[width=\columnwidth]{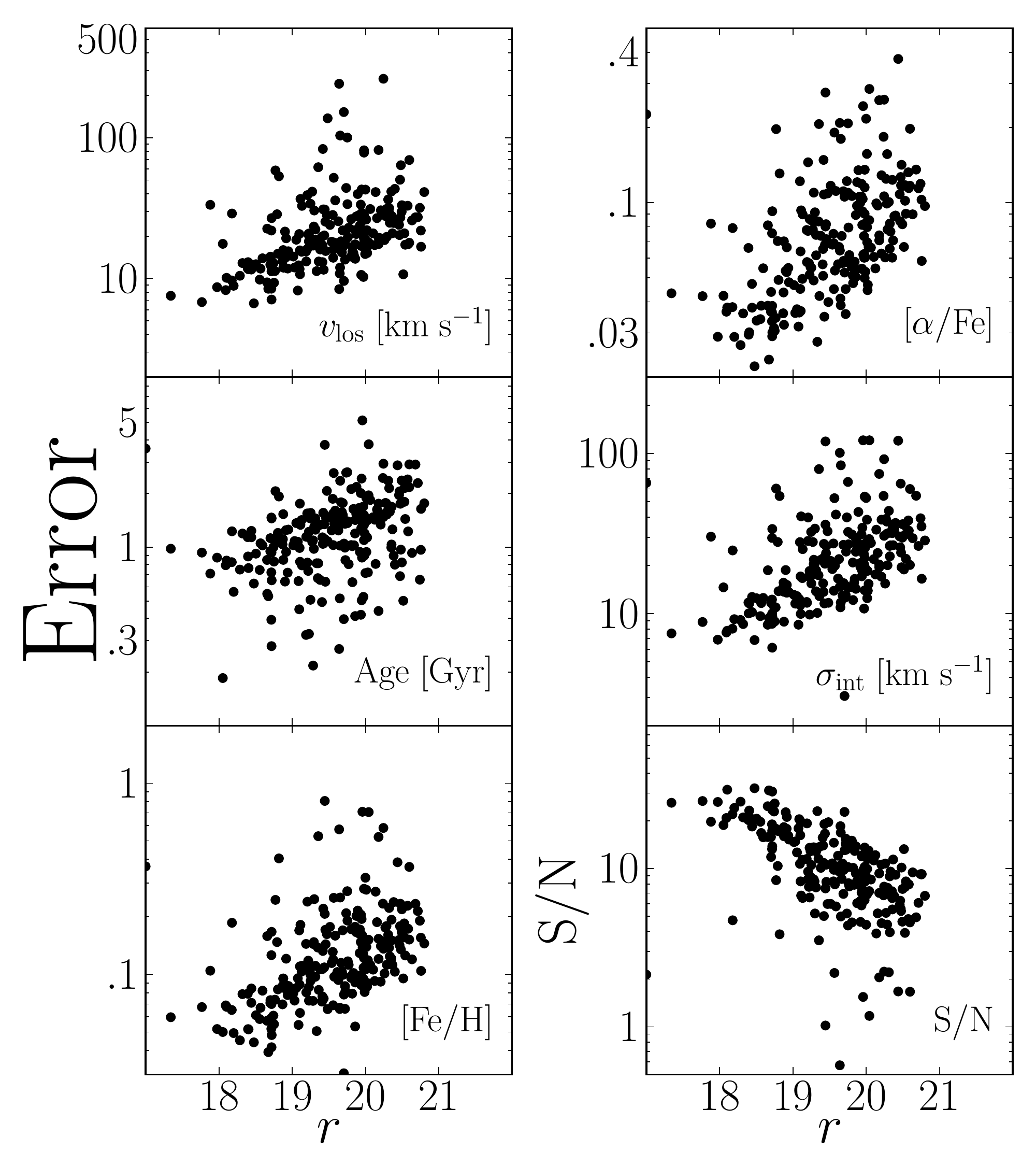}
\caption{Errors in the five galactic parameters as a function of r-band magnitude and median pixel S/N as a function of r-band magnitude (each panel is labeled in the bottom right corner).}
\label{A267MagPlots}
\end{figure}

As a first application of our model, we fit new data of the cluster A267.
In order to measure the LSF of M2FS, we first fit a set of twilight spectra.
In doing so, we estimate the posterior probability distribution of each of the 3 $h_n$ parameters (see \S\ref{Spectral Model}).
Because we observed one twilight spectrum for each of the 256 fibers of M2FS, we quantify the posterior PDFs of the LSF for each fiber independently.
Then, when fitting each of the science spectra, we sample the PDFs of the LSF that corresponds to the fiber that the science spectrum was observed.
This technique quantifies the LSF and allows our fitting routine to break the degeneracy between the LSF and $\disp$; furthermore, it also naturally propagates the uncertainty in each of the $h_n$ parameters into $\disp$ for each galaxy spectrum.
Fig. \ref{AllLSFs} shows the resolving power $R=\Delta\lambda/\lambda$ for all fibers used in this analysis.
For each fiber we used the central 68\% of the LSF covered by the PDFs determined by \textsc{multinest} to calculate $R$, which we plotted in Fig. \ref{AllLSFs}.
The two colors in Fig. \ref{AllLSFs} correspond to the separate spectrographs that are used in M2FS.
There is a clear dichotomy between the spectrographs with the ``blue" channel giving a systematically higher resolution; nonetheless, $R$ is roughly centered around the theoretical resolving power of M2FS at the low-resolution configuration of $\sim2200$.

After fitting all twilight spectra, we then fit the sky-subtracted science spectra using the technique described in \S\ref{Analysis of Spectra} above.
Table \ref{A267Results} shows the results of these fits for the galactic parameters.
The parameter estimations are multidimensional posterior PDFs.
Therefore, in Table \ref{A267Results} we give the mean value of the marginal PDFs for each parameter as well as the width of these distributions (central 68\%) shown as an error.

Fig. \ref{A267BestFits} shows a series of sky-subtracted A267 spectra plotted in blue.
Over plotted in red is the range of model fits covering the central 68\% of the posterior probability distribution.
Essentially, the red regions (thick red lines) shows the width of the posterior PDF converted into a spectrum.
The bottom panel of each plot shows the residuals scaled by the variance in each pixel.
Here we are only showing the residuals for the spectrum corresponding to the set of best fit parameters.
The residuals scaled by the variance in each pixel is given by
\begin{equation}
\label{delta residuals}
\delta(\lambda)=\frac{S(\lambda)-M(\lambda)}{\sqrt{\mathrm{Var}[S(\lambda)]}}
\end{equation}
where $S(\lambda)$ is the sky-subtracted science spectrum, $M(\lambda)$ is the best fit model, and $\mathrm{Var}[S(\lambda)]$ is the measured variance in the science spectrum.
Also shown in each plot are the best fit galactic parameters along with their uncertainties, which are equal to the widths of their 1D posterior PDFs.
To show the effectiveness of our model as a function of S/N, we arranged the plots with high median S/N per pixel in the top two panels ($\mathrm{S/N}\sim30$) to mid-level S/N in the middle ($\sim15$) to low S/N in the bottom ($\sim2$).
Furthermore, the set of plots in the left column are for spectra with a high probability of membership to A267, while the spectra on the right are foreground and background galaxies.

In each of the plots in Fig. \ref{A267BestFits}, we show the set of values for the galactic parameters corresponding to the best fit (highest likelihood) of the model to the data along with their uncertainties.
We display the multi-dimensional posterior PDF of the five physical galactic parameters in Fig. \ref{A267PosteriorPart}.
Here, one can more easily see the effectiveness of our model to constrain the physical parameters of interest.
For the 2D marginal PDFs, we once again show the $1\sigma$ and $2\sigma$ contours as the dark and lighter shaded regions, respectively, in each panel.
Most of the PDFs in Fig. \ref{A267PosteriorPart} are Gaussian in shape and therefore can be easily quantified by a mean (or a highest likelihood value) and a variance; however, some parameters (i.e. Age in Fig. \ref{A267PosteriorPart}) have some non-Gaussian features.
Because of this non-Gaussianity, it is better to describe the best fit parameters by a PDF instead of a single value and a variance.
Having said that, we can still see in Fig. \ref{A267PosteriorPart} that the highest likelihood parameter values still estimate the mean of the posterior PDFs effectively and the variance in these values still gives a good approximation of the width of these distributions.

Fig. \ref{A267MagPlots} shows the error for each of the five galactic parameters labeled in the top right of each panel and median pixel S/N as a function of r-band magnitude.
We notice usual behavior for our observations: for fainter objects, median pixel S/N decreases while errors in measured quantities increase.
Parameter estimates of A267 have median random errors of $\sigma_{v_\mathrm{los}}=20\ \mathrm{km\ s^{-1}}$, $\sigma_\age=1.2\ \mathrm{Gyr}$, $\sigma_\feh=0.11\ \mathrm{dex}$, $\sigma_{[\alpha/\mathrm{Fe}]}=0.07\ \mathrm{dex}$, and $\sigma_{\sigma_\mathrm{int}}=20\ \mathrm{km\ s^{-1}}$.

All raw spectra, our spectral fits, and all posteriors attributed to these fits are fully available online at the Zenodo database: https://doi.org/10.5281/zenodo.831784.

\subsection{Comparison to previous redshift results}
\label{zComp}

\begin{figure}
\centering
\includegraphics[width=\columnwidth]{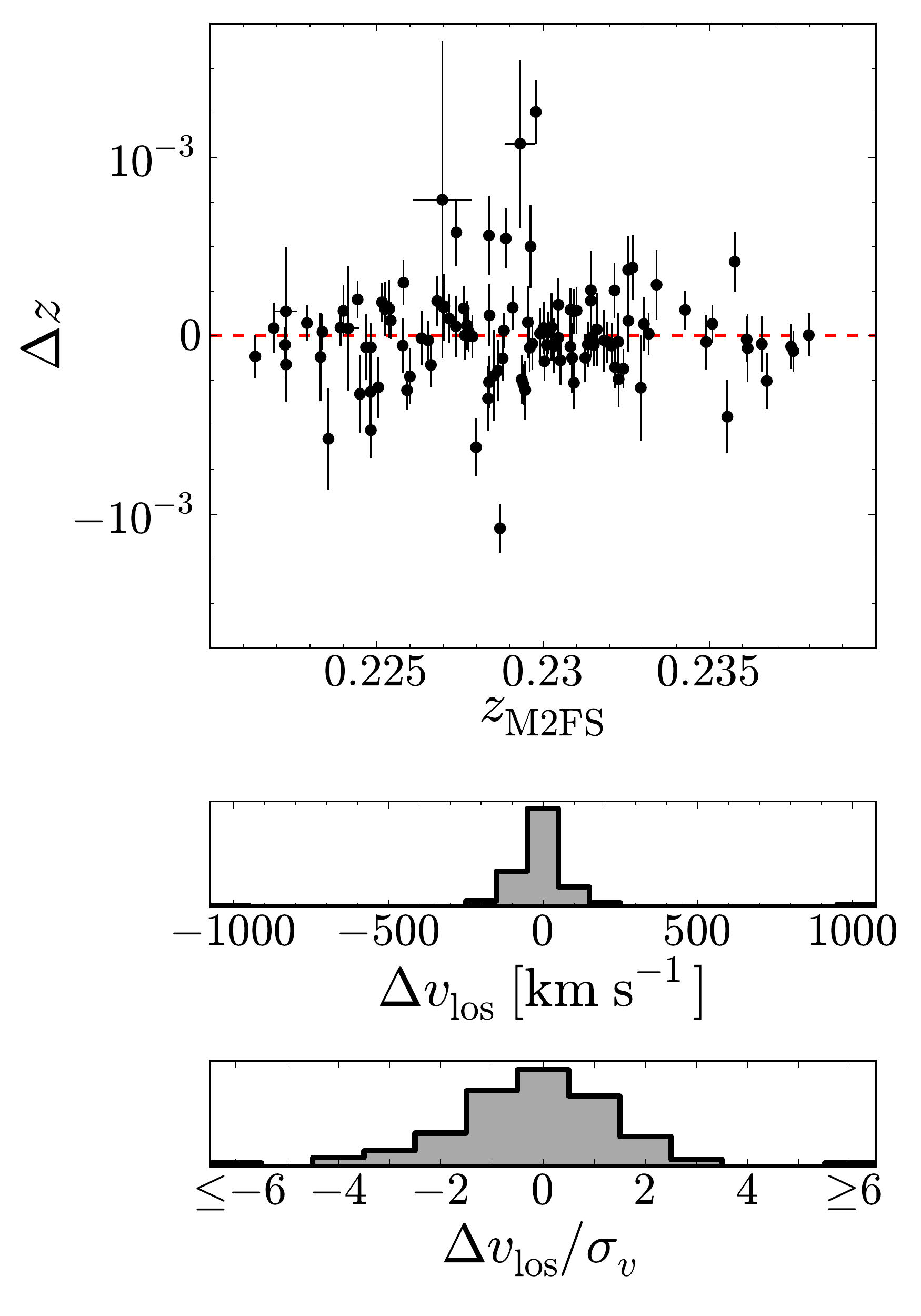}
\caption{Comparing the measured redshifts of each galaxy from our analysis $z_\mathrm{M2FS}$ to previously published results in \citet{Rines2013}.  For clarity in the top panel we only show galaxies with measured redshifts around that of A267 $z\sim0.23$; however, the bottom two panels show the distribution for all overlap observations with \citet{Rines2013}. The bottom two panels shows the distribution in differences of redshifts, the bottom most is scaled by the combined uncertainties in the two measurements.}
\label{A267GellerCompare}
\end{figure}

In order to discuss the accuracy of our A267 fits, we compare our redshifts to those measured previously by \citet{Rines2013}.
In their paper they measured redshifts for over 22,000 galaxies from The Hectospec Cluster Survey (HeCS), they cite 226 galaxy members to A267, and we re-observe 114 of those.
In Fig. \ref{A267GellerCompare} we compare our measured redshifts ($z_\mathrm{M2FS}$) to theirs.
In the top panel of Fig. \ref{A267GellerCompare}, for added clarity, we only show galaxies that are approximately at the redshift of A267 ($z\sim0.23$); on the other hand, the histograms in the bottom two panels show the distribution for all 196 repeat observed galaxies (separation $<5\times10^{-5}$deg).
The top histogram panel shows the difference in the measured line-of-sight velocity $\Delta v_\mathrm{los}$, while the bottom most panel shows this difference scaled by the combined uncertainties in the measured redshifts $\sigma_v$.
The histograms show that our redshift measurements are in good agreement with those measured by \citet{Rines2013}.

\section{Conclusions}
\label{conclusions}
We have introduced a new model for fitting galaxy spectra using a Bayesian approach and integrated light spectra.
We chose to produce a new integrated light model for a few important reasons, which we highlight in the paper.
The main reason is that we wish to implement this modeling in the Bayesian statistical framework offered by MultiNest, which allows us to fully quantify the covariances of all free parameters.
Furthermore, our new model gives us the flexibility to alter any aspect of the model from pre-calculated isochrones, to choices of synthetic spectral libraries, to complexity of stellar populations, which would be difficult to implement in the previous population synthesis techniques.
In \S\ref{External Tests}, we showed that this model is able to adequately reproduce the results of previous stellar populations fits to A496 spectra, while increasing flexibility for measuring the internal velocity dispersion of the stellar population.
Lastly, this model robustly incorporates a wavelength dependence fit for the line-spread-function without the use of Hermite-Gaussian polynomials, which are typically used.

We outlined the process we used to generate an integrated light spectral library from a pre-calculated database of isochrones (Dartmouth Isochrones) and a library of synthetic stellar spectra (Phoenix Spectral Library).
For this calculation, we assumed a Chabrier log-normal IMF with fixed scaling parameters, but the choice of IMF can be changed to incorporate different stellar evolution theories as well as allowing the parameters or the model be free.
Furthermore, the choice of isochrones and stellar library can vary and one could use a library of real stellar spectra instead.
We then discussed the model used to fit the galaxy operations and how we fit this model using the Bayesian nested sampling algorithm MultiNest.

In order to test the statistical power of the model, we generated and fit a mock catalog of galaxy spectra thus quantifying the accuracy of the model.
This showed that for increasing S/N, the model performs better; however, even for low S/N$\sim5$, we are still able to reproduce the input galactic parameters with some level of precision.
Furthermore, some of the galactic parameters are more easily estimated at lower S/N.
For example, the velocity of the galaxy $\vlos$ can be estimated from our model with a high degree of certainty over the full range of S/N tested with the mock catalogs; however, we achieved similar precision for the galactic age parameter at only high S/N values.

Following the analysis of the mock catalog, we applied the integrated light spectral model to new spectral data acquired from M2FS on the Clay Magellan Telescope.
We fit these spectra and estimated the posterior probability distribution for five galactic parameters: $\vlos$, $\age$, $\feh$\, $\afe$, and $\disp$.
We compared the estimates of $\vlos$ to previously published measurements from \citet{Rines2013}, which shows much agreement between the two measured redshifts.

In a companion paper, we will use our spectroscopic measurements to model the internal dynamics and galaxy populations of A267.
In the companion paper we will apply a multi-population Dynamical Jeans Analysis.
This model will simultaneously fit the dark matter and light distributions within the cluster while identifying contamination galaxies, substructure within the cluster environment and any overall cluster rotation.

\section*{Acknowledgements}

We thank Margaret Geller and Ken Rines for helpful discussions that improved the quality of this work.
We thank Charlie Conroy and Ben Johnson for useful discussions on incorporating wavelength dependence into the line-spread-function of the model.
E.T. and M.G.W. are supported by National Science Foundation grants AST-1313045 and AST-1412999.  M.M. is supported by NSF grant AST-1312997.  E.W.O. is supported by NSF grant AST-1313006.
We would also like to thank the anonymous referee for their helpful comments in improving this paper.

\bibliography{bib}{}
\bibliographystyle{aasjournal}
\label{lastpage}
\end{document}